\documentclass[preprintnumbers,showpacs,amsmath,amssymb,a4paper]{revtex4} 
\usepackage[]{epsfig} 
 
\newcommand{\Teff}{T_{\rm eff}} 
\newcommand{\Xin}{X^\infty} 
\newcommand{\tw}{\ensuremath{t_\mathrm{w}}} 
\newcommand{\dt}{\ensuremath{\Delta t}} 
 
\newcommand{\upd}{{\rm d}} 
\newcommand{\me}{\mathrm{e}} 
\newcommand{\In}{\mathrm{I}} 
\newcommand{\Hn}{\mathrm{H}} 
\newcommand{\erfi}{\mathrm{Erfi}} 
 
\newcommand{\cors}[1][\mathrm{s}]{C_{#1}} 
\newcommand{\ress}[1][\mathrm{s}]{R_{#1}} 
\newcommand{\chis}[1][\mathrm{s}]{\chi_{#1}} 
\newcommand{\Xs}[1][\mathrm{s}]{X_{#1}}

\newcommand{\corm}{C_{\mathrm{m}}} 
\newcommand{\resm}{R_{\mathrm{m}}} 
\newcommand{\chim}{\chi_{\mathrm{m}}} 
\newcommand{\Xm}{X_{\mathrm{m}}} 

\newcommand{\cord}[1][\mathrm{d}]{C_{#1}} 
\newcommand{\resd}[1][\mathrm{d}]{R_{#1}} 
\newcommand{\chid}[1][\mathrm{d}]{\chi_{#1}} 
\newcommand{\Xd}[1][\mathrm{d}]{X_{#1}} 

\newcommand{\core}{C_{\mathrm{e}}} 
\newcommand{\rese}{R_{\mathrm{e}}} 
\newcommand{\chie}{\chi_{\mathrm{e}}} 
\newcommand{\Xe}{X_{\mathrm{e}}} 
 
\newcommand{\lorenz}{\mathrm{L}} 
\newcommand{\power}{\mathrm{P}} 
 
\begin{document} 
 
\title{Fluctuation-dissipation relations in the non-equilibrium critical 
dynamics of Ising models} 
 
\author{Peter Mayer$^1$, Ludovic Berthier$^{2,3}$, Juan
P. Garrahan$^2$ and Peter Sollich$^1$}

\affiliation{$^1$Department of Mathematics, King's College, Strand,
London, WC2R 2LS, UK \\ $^2$Theoretical Physics, University of Oxford,
1 Keble Road, Oxford, OX1 3NP, UK \\ $^3$Laboratoire des Verres,
Universit\'e Montpellier II, 34095 Montpellier, France}
 
\date{\today} 
 
\begin{abstract} 
We investigate the relation between two-time, multi-spin, correlation
and response functions in the non-equilibrium critical dynamics of
Ising models in $d=1$ and $d=2$ spatial dimensions.  In these
non-equilibrium situations, the fluctuation-dissipation theorem (FDT)
is not satisfied. We find FDT `violations' qualitatively similar to
those reported in various glassy materials, but quantitatively
dependent on the chosen observable, in contrast to the results
obtained in infinite-range glass models.  Nevertheless, all FDT
violations can be understood by considering separately the
contributions from large wavevectors, which are at quasi-equilibrium
and obey FDT, and from small wavevectors where a generalized FDT holds
with a non-trivial fluctuation-dissipation ratio $\Xin$.  In $d=1$, we
get $\Xin = \frac{1}{2}$ for spin observables, which measure the
orientation of domains, while $\Xin = 0$ for observables that are
sensitive to the domain-wall motion. Numerical simulations in $d=2$
reveal a unique $\Xin \simeq 0.34$ for all observables.  Measurement
protocols for $\Xin$ are discussed in detail.  Our results suggest
that the definition of an effective temperature $\Teff = T / \Xin$ for
large length scales is generically possible in non-equilibrium
critical dynamics.
\end{abstract} 
 
 
\maketitle 
 
Since the analytical solution of the non-equilibrium dynamics of the
spherical $p$-spin model in its low temperature phase~\cite{cuku},
many studies have focused on the properties of two-time
non-equilibrium correlation and response functions, and the
relationship between them~\cite{review_aging,leticiazouches}.  In this
paper, we report on analytical and numerical investigations of several
two-time, multi-spin, correlation and response functions in the
non-equilibrium critical dynamics of Ising models in $d=1$ and $d=2$
spatial dimensions.
 
Our work is motivated by the following observations. Multi-point
dynamical functions are standard objects in equilibrium statistical
mechanics which reveal microscopic information related to
experimentally observable quantities.  In non-equilibrated systems,
however, the equilibrium relation between response and correlation,
i.e.\ the fluctuation-dissipation theorem (FDT), is not
satisfied. This evident observation became important when it was
realized that in the $p$-spin model~\cite{cuku} and more generally in
infinite-range glass models, a {\em generalized} FDT can be
formulated~\cite{review_aging,leticiazouches,cuku2,mefr}.  This
amounts to the introduction of a fluctuation-dissipation ratio $X$ or,
alternatively, of an effective temperature $\Teff=T/X$ for the slow,
non-equilibrated modes of the system~\cite{cukupe}.  The properties of
$X$ and $\Teff$ have attracted much interest since they suggest that a
generalized statistical mechanics can be devised to deal with a broad
class of non-equilibrium phenomena.
 
The generalized FDT is exact for infinite-range glass models only. It
is, however, tempting to apply the same concepts in other contexts
such as glassy systems with finite interaction range, as observed
experimentally or simulated numerically.  A further step is made when
those ideas are transferred to other physical situations such as
domain growth processes~\cite{FDTferro1,a1} in non-disordered systems
or the rheology of soft glassy materials~\cite{SGR_ref,BBK}. Although one
does not expect the results for infinite-range glass models to apply
exactly in all these non-equilibrium situations, it is worthwhile to
understand analogies and differences, and thus to push these ideas as
far as possible. This is the philosophy of our paper where
non-equilibrium dynamics at criticality is analyzed along the lines
described above.  Our results suggest that the concept of a
generalized FDT is indeed useful at criticality, and we describe in
detail the form it takes as compared to the results obtained in
infinite-range glass models.
 
The manuscript is organized as follows.  The first section below
reviews the results obtained for correlation and response functions in
ferromagnets and delineates the scope of the paper.  In
Sec.~\ref{section1d}, the 1$d$ Ising model is studied analytically at
$T_c=0$. In Sec.~\ref{section2d}, numerical results for the 2$d$ Ising
model at $T_c$ are presented. A summary and a physical discussion of
the results can be found in Sec.~\ref{summary}.
 
\section{FDT and ferromagnets} 
 
\subsection{Correlation and response functions} 
 
Pure ferromagnets are generally not described as glassy materials,
which are loosely defined as systems with large relaxation times.
However, if a ferromagnet initially prepared at high temperature is
suddenly quenched to its low temperature ferromagnetic phase, its
equilibration time diverges with system size~\cite{review_alan}.  This
is true also when the quench is performed precisely to the critical
point, $T=T_c$.  In both cases the system remains, in the
thermodynamic limit, forever in a non-equilibrated, non-stationary
state: it exhibits aging.  Therefore, one can study physical
situations in pure ferromagnets that are reminiscent of aging
phenomena observed e.g.\ in spin glasses, polymers or colloids.  One
is then led to ask if the tools used in the glass literature are also
useful to describe this type of non-equilibrium situation.
 
These tools include, in particular, two-time correlation and response
functions.  Consider two physical observables $A(t)$ and $B(t)$.
Their connected cross-correlation is defined by
\begin{equation} 
  C(t,\tw) = \langle A(t) B(\tw) \rangle - \langle A(t) \rangle 
  \langle B(\tw) \rangle,  
  \label{corrAB} 
\end{equation} 
while the conjugate response function is given by 
\begin{equation} 
  R(t,\tw) = T \left. \frac{\delta \langle A(t) \rangle}{\delta  
  h_B (\tw)} \right|_{h_B=0}.
  \label{respAB} 
\end{equation} 
Here $h_B$ is the thermodynamically conjugate field to the observable
$B$; for later convenience we scale the response by $T$.  Numerically
or experimentally, it is often more convenient to measure the
integrated response function, or susceptibility,
\begin{equation} 
  \chi(t,\tw) = \int_{\tw}^t \upd \tau \, R(t,\tau),
  \label{chiAB} 
\end{equation} 
which gives the response to a small constant field $h_B$ switched on
at the `waiting time' $\tw$.
 
At equilibrium, correlation and response are time-translation 
invariant and related by the FDT 
\begin{equation} 
R(t,\tw) = \frac{\partial}{\partial \tw} C(t,\tw). 
\end{equation} 
In that case, $\langle \ldots \rangle$ in 
Eqs.~(\ref{corrAB},\ref{respAB}) stands for the 
usual ensemble average.  
If one follows instead the dynamics of the system after a sudden quench,  
the system is out of equilibrium and  
neither time-translation invariance 
nor FDT are satisfied. Then $\langle \ldots \rangle$ is to be read as 
an average over initial conditions and any stochasticity in the dynamics. 
In infinite-range glass models, 
a generalized FDT is satisfied in the aging dynamics.  
The generalization amounts 
to the introduction of a fluctuation-dissipation ratio  
(FDR), $X(t,\tw)$, through  
\begin{equation} 
  -\frac{\partial }{\partial \tw}\chi(t,\tw) = 
  R(t,\tw) = X(t,\tw)  
  \frac{\partial }{\partial \tw} C(t,\tw). 
  \label{FDTAB} 
\end{equation} 
This definition becomes non-trivial because in the limit of 
long times $t$ and $\tw$ the FDR reduces to  
a function of a single variable only, namely the correlation function, 
\begin{equation} 
  X(t,\tw) \to X(C(t,\tw)),
  \label{FDRAB} 
\end{equation} 
where by a an abuse of notation we retain the same symbol for the FDR
and its long-time limit. As in equilibrium, response and correlation
are then no longer independent quantities, although their relationship
is now more complex.
  
It is now standard to study this generalized FDT via the parametric
representation, or `FD plot', of the susceptibility $\chi$ as a function
of the correlation $C$~\cite{cuku2}.  At equilibrium, one has
$\chi(t,\tw) = C(t,t) - C(t,\tw)$.  Hence, a plot of $\chi(t,\tw)$
against $C(t,t) - C(t,\tw)$ gives a simple straight line of slope $1$:
this is the equilibrium FD plot.  Out of equilibrium,
Eq.~(\ref{FDRAB}) implies
\begin{equation} 
  \chi(t,\tw) = \int_{C(t,\tw)}^{C(t,t)}  \upd x \, 
  X(x). 
\end{equation} 
Therefore, when Eq.~(\ref{FDRAB}) holds, the FDR can be obtained
directly from the slope of the FD plot, which is $X(C)$.  Otherwise,
from Eq.~(\ref{FDTAB}) a plot of $\chi(t,\tw)$ against $C(t,t) -
C(t,\tw)$, with $t$ fixed and $\tw$ the curve parameter, will still have
slope $X(t,\tw)$. Since the amplitude of correlation and response
functions can diverge or converge to zero for $t\to\infty$ (see below)
it can be useful to use normalized quantities, plotting
$\tilde{\chi}(t,\tw) \equiv \chi(t,\tw)/C(t,t)$ versus
$1-\tilde{C}(t,\tw)$ where $\tilde{C}(t,\tw)=C(t,\tw)/C(t,t)$.  Since
the normalization factors are independent of $\tw$, the slope of the
FD plot is then still given by $X$. The normalization issue is less
important when presenting numerical data, which are by construction
obtained in a restricted time window where the amplitudes of the
dynamical quantities typically change only slowly.
 
Appealingly, the FDR can also be interpreted as defining an effective
temperature, $\Teff(t,\tw) \equiv T/X(t,\tw)$, replacing the
equilibrium temperature $T$ by an equivalent quantity out of
equilibrium.  Moreover, it is a general result that for the case
$A=B$, where one considers the autocorrelation of $A$ and the
associated response, $\Teff$ is the temperature measured by a
thermometer coupled to the observable $A$ at the
appropriate timescale~\cite{cukupe}.  
As a direct corollary, this effective
temperature then satisfies the zeroth law of thermodynamics.  Clearly,
however, the introduction of an effective temperature is of
thermodynamic interest only if $\Teff$ is actually independent of the
observables $A$ and $B$ under consideration.  This is indeed true for
infinite-range glass models~\cite{cukupe}, implying that although the
system is out of equilibrium it can still be described in
thermodynamic terms, at the moderate cost of introducing one extra
parameter, namely the effective temperature~\cite{theo}.  Beyond
infinite-range glass models, the observable-dependence of the
effective temperature remains largely an open question but has been
discussed in detail in the context of trap models~\cite{petersuzanne}
and in a realistic numerical model of a supercooled
liquid~\cite{BeBa}.

\subsection{Ferromagnets at low temperature} 
 
For ferromagnets, two-time dynamical functions have been
studied both for a quench to the low temperature phase and to
criticality, with most work to date focused on the first situation.
 
In the low temperature phase, the evolution of the system consists in
the growth of ordered domains, with a typical domain size $\ell(t)$.
Two-time quantities that have been thoroughly studied are the spin
autocorrelation function, $\cors(t,\tw) = \langle s_i(t) s_i(\tw)
\rangle$, where $s_i(t)$ is the value of the spin at site $i$ at time
$t$, and the conjugate response function, $\ress(t,\tw) = -
(\partial/\partial\tw)\chi_{\rm s}(t,\tw) = \delta \langle s_i(t)
\rangle / \delta h_i(\tw)$, where $h_i$ is the magnetic field at site
$i$.  In this case, the connected and disconnected correlations
coincide since $\langle s_i (t) \rangle = 0$ at all stages of the
coarsening process. From the analytical solution of solvable
models~\cite{cukupa,FDTferro2,horner,FDTferro13,italians1} and the
simulation of more realistic
situations~\cite{FDTferro1,FDTferro4,a1,FDTferro3,italians2}, the
behavior of these two quantities is now well known.  For small time
differences, $\dt=t-\tw \ll \tw$, time-translation invariant behavior
is observed, $\cors(t,\tw) \approx \cors(\dt)$, $\chis(t,\tw) \approx
\chis(\dt)$, and the FDT $\chis(t,\tw) \approx
\cors(t,t)-\cors(t,\tw)$ is obeyed.  This first regime is due to
thermal fluctuations in the bulk of the domains, which are essentially
equilibrium fluctuations.  For larger time separations, $\dt \gg \tw$,
the fluctuations of the interfaces dominate the dynamical behavior.
The reasonable hypothesis that coarsening is a self-similar process,
in the sense that for large times all dynamical functions depend on
time only through the typical domain size $\ell$, implies the scaling
form $\cors(t,\tw) \approx \cors (\ell(t)/\ell(\tw))$ for the
correlation function.  The contribution of the interfaces to the
response function can be estimated~\cite{a1} as a sum over all
wavevectors, $\chis(t,\tw) \approx \int_{1/\ell(\tw)}^{1/a} \upd^d
{\mathbf k}\, k^{-2} \ell^{-1}(t)$.  This expression results from the
fact that the response at time $\tw$ is dominated by large wavevectors,
$k \ell(\tw) \gg 1$, each wavevector ${\mathbf k}$ giving a
contribution of the order of $k^{-2}$.  The factor $\ell^{-1}(t)$
represents the density of domain walls, and $a$ is a UV cutoff given
by the lattice spacing.  This reasoning implies for the long-time
contribution to the susceptibility the scaling form $\chis(t,\tw)
\approx f(\ell(\tw)) 
\, \chis  (\ell(t)/\ell(\tw))$, where the function
$f(x)$ depends on the dimensionality $d$ of space and is given by
$f(x) = 1/x$ for $d>2$, $f(x) = (\ln x)/x$ for $d=2$ and $f(x) =$
const for $d=1$. This scaling function has recently been revisited in
Refs.~\cite{italians2,henkel}, with particular attention to the case
$d=2$~\cite{coco,henkel2}.
 
From the above arguments, and for $d>1$, the parametric plot of
$\chis(t,\tw)$ versus $\cors(t,\tw)-\cors(t,\tw)$ consists of an
initial equilibrium part followed by an essentially horizontal
section. In the latter the correlation function decays due to
interface motion, while the response function hardly changes because
any contribution from the interfaces is suppressed by the $f(\ell(\tw))$
prefactor.  If a limiting FDR, $\Xs^\infty$, is defined through
\begin{equation} 
  \Xs^\infty = \lim_{\tw \to \infty} \lim_{t \to \infty} \Xs(t,\tw), 
\end{equation} 
then it follows that $\Xs^\infty = 0$ for $d>1$ in coarsening
processes.  For $d=1$, on the other hand, both $\chis$ and $\cors$ are
scaling functions of $\ell(t)/\ell(\tw)$ and the parametric plot
assumes no simple shape, implying that $\Xs^\infty$ could be any
finite number.  This is confirmed by the analytical solution of the
dynamics of the Ising chain at $T=0$ which shows that $\Xs^\infty =
\frac{1}{2}$~\cite{golu1,zaza}.  The factor $\frac{1}{2}$ was first
derived in Ref.~\cite{hilhorst}.

\subsection{Ferromagnets at the critical point} 
\label{ferrocrit} 
 
The non-trivial value of $\Xs^\infty$ for the Ising chain was
interpreted using the fact that in $d=1$, the ordering temperature
$T=0$ coincides with the critical point, $T_c=0$~\cite{golu1}.  It was
then suggested that a non-trivial $\Xs^\infty$ could be a generic
feature of critical points~\cite{golu2}.  This is physically
reasonable, since the whole argument for $\Xs^\infty = 0$ in
coarsening processes relies on the separation between bulk and
interfaces; this is no longer valid at the critical point where the
bulk has the well-known self-similar structure of ferromagnets at
criticality.
 
From analytical and numerical studies, the behavior of two-time
single-spin dynamical quantities is again well understood, as
briefly reviewed in Ref.~\cite{golu3}.  Physically, the
non-equilibrium dynamics following a quench to the critical point
consists in the growth of the dynamical correlation length, $\xi(t)
\approx t^{1/z}$, where $z$ is the dynamical critical
exponent~\cite{hoha}.  Critical fluctuations of large wavevectors, $k
\xi(t) \gg 1$, are almost equilibrated, while those with small
wavevectors, $k \xi(t) \ll 1$, still retain their non-equilibrium
initial condition.  This separation leads to the scaling forms
$\cors(t,\tw) \approx {\dt}^{- 2 \beta / \nu z}
\cors(\xi(t)/\xi(\tw))$ and $\chis(t,\tw) \approx {\dt}^{- 2 \beta /
\nu z} \chis(\xi(t)/\xi(\tw))$, where $\beta$ and $\nu$ are the
standard critical exponents.  This can be interpreted as follows. For
short time differences $\dt \ll \tw$, equilibrated fluctuations with
large $k$ dominate and dynamical functions assume their standard
equilibrium power-law decay.  The dynamics at large time separation
$\dt \gg \tw$, on the other hand, is dominated by the growth of the
dynamic correlation length and leads to the $\xi(t)/\xi(\tw)$
scaling. This in turn implies that, beyond the initial equilibrium
part, the FD plot will again assume a non-trivial shape, as in the
Ising chain.  The striking similarity of these results with the aging
dynamics of finite dimensional spin glasses was noted in
Refs.~\cite{golu2,a3}.
 
The reasoning above confirms that at criticality $\Xs^\infty$ can 
take any finite value, in contrast with the $\Xs^\infty=0$ obtained in 
the low temperature phase.  It was further argued that $\Xs^\infty$ 
should be a new universal quantity at criticality~\cite{golu2}.  As 
such, it can be computed using standard renormalization group 
procedures, and this program has recently been started for various 
models~\cite{gambas1,gambas2,gambas3}. The value of $\Xs^\infty$ is 
known exactly for the Ising chain~\cite{golu1,zaza,hilhorst}, where 
$\Xs^\infty=\frac{1}{2}$, for the spherical ferromagnetic 
model~\cite{golu2}, where $\Xs^\infty=\frac{1}{2}$ for $d \ge 4$ and 
$\Xs^\infty=\frac{1}{3}$ for $d=3$, and for the Gaussian 
model~\cite{cukupa} where $\Xs^\infty=\frac{1}{2}$.  An estimate is known 
for model A at second order in $4-d$~\cite{gambas1}, to first order in 
$\sqrt{4-d}$ for the diluted Ising model~\cite{gambas2}, and to first 
order in $4-d$ in model C~\cite{gambas3}.

\subsection{Motivations for this work} 
\label{ferromotiv} 
 
This short review of known results in the non-equilibrium dynamics of
pure ferromagnets shows that much research has been done on the
subject. So, why another paper?
 
First of all, the relevance of the notion of an effective temperature
at criticality can be questioned because the FD plots for the spin
dynamic functions do not assume a simple linear shape with a
well-defined slope, as happens in the low temperature phase.  This is
related to the fact that at low temperatures the decay of correlation
functions occurs on two well-separated time scales. Each has its own
associated effective temperature, a fact reminiscent of the physics of
structural glasses.  At criticality, on the other hand, one has a
continuum of time scales associated with different wavevectors, $t(k)
\sim k^{-z}$. Moreover, for finite $k$ the equilibration time is
finite, meaning that the number of modes that are still out of
equilibrium decreases as time increases.  This suggests that an
effective temperature could be relevant only when considering the $k
\to 0$ limit, a point which our analysis will clarify.
 
Second, we mentioned the important issue of the observable-dependence
of a generalized FDT. This issue remains completely open since the
studies cited above focused exclusively on a single FD relation, for
the spin autocorrelation and associated response. In order to get a
more complete theoretical understanding it is crucial to understand if
other observables give the same results, and if not, how they are
related.
 
A third motivation for the study of higher order correlation functions
comes from the observation that the dynamics of coarsening models is
dominated by the motion of topological defects. For Ising models,
these are domain walls, the local density of which is given by the
`defect' observable $s_i (t) s_j(t)$, where $(i,j)$ are nearest
neighbors. Defect dynamical functions have recently been studied in
the context of kinetically constrained Ising models~\cite{juanpe1},
and the FD relations that arose showed interesting and unexpected
features.
 
For an Ising model, there are at least four `natural' FD relations,
involving respectively the spin autocorrelation, $\cors(t,\tw)$, the
magnetization density $m(t)$ correlation, $\corm(t,\tw) = \langle m(t)
m(\tw) \rangle$, the defect autocorrelation, $\cord(t,\tw) = \langle
s_i(t) s_j(t) s_i(\tw) s_j(\tw) \rangle - \langle s_i(t) s_j(t)
\rangle \langle s_i(\tw) s_j(\tw) \rangle$ with $(i,j)$ nearest
neighbors, and the energy density $e(t)$ correlation, $\core(t,\tw) =
\langle e(t) e(\tw) \rangle - \langle e(t) \rangle \langle e(\tw)
\rangle$.  Note again that connected and disconnected correlation
functions coincide for the magnetization; this is not the case for
$\cord(t,\tw)$ and $\core(t,\tw)$.  In the $1d$ case, we will also
investigate two-time functions which smoothly interpolate between
incoherent, local functions (spin or defect) and coherent, global ones
(magnetization, energy), and discuss the case of correlation functions
of higher order.  In the 2$d$ case, we will stick to the four
quantities listed above.

\section{The 1$\boldsymbol{d}$ Ising model} 
\label{section1d} 
 
In this section we study the non-equilibrium dynamics in the
Glauber-Ising chain with Hamiltonian
\begin{equation} 
  \mathcal{H} = - \sum_i s_i \, s_{i+1}, 
\end{equation} 
where the $s_i$ ($i=1,\ldots N$) are $N$ Ising spins subject to
periodic boundary conditions. Glauber dynamics consists in each spin
$s_i$ flipping with rate $\frac{1}{2}[1-\frac{1}{2}\gamma
s_i(s_{i-1}+s_{i+1})]$, where $\gamma = \tanh(2/T)$. We focus on the
evolution of arbitrary two-time spin and defect correlation and
response functions in the thermodynamic limit $N \rightarrow \infty$,
after a quench from equilibrium at $T=\infty$ to $T \rightarrow 0$. As
explained above, although a variety of aspects of the associated
coarsening dynamics have already been
studied~\cite{review_alan,alan1d}, results on the non-equilibrium FDT
violation so far are restricted to the spin autocorrelation and
response functions \cite{golu1,zaza}.  In Sec.~\ref{general} we
introduce the more general class of spin and defect observables we
investigate. We briefly present the main result of our method to
derive multi-spin two-time correlation and response functions, as
developed in Ref.~\cite{peterpeter2}, and summarize the approach used
to extract from this the quantities of interest here. Our results for
spin dynamical functions are then given in Sec.~\ref{2spin1d} ---a
preliminary account of which has appeared in the conference
proceedings~\cite{peterpeter1}--- and for defect functions in
Sec.~\ref{4spin1d}. In Sec.~\ref{physdis} we discuss the physical
aspects of our results for the 1$d$ Ising model.

\subsection{General strategy for the calculations} 
\label{general} 
 
We consider the following spin and defect observables $O_\mathrm{s}$
and $O_\mathrm{d}$
\begin{equation} 
  O_\mathrm{s}=\sum_i \epsilon_i s_i  
  \quad \mbox {and} \quad  
  O_\mathrm{d}=\sum_i \epsilon_i s_i s_{i+1}. 
\label{equ:observables} 
\end{equation} 
In both cases the $\epsilon_i$ are quenched random variables with zero
mean $[\epsilon_i]=0$ and translation invariant covariances
$q_{i-j}=[\epsilon_i\epsilon_j]$; here $[\,\cdot\,]$ denotes the
average over the distribution of $\epsilon$. Without loss of
generality we set $q_0=1$.  We define the corresponding connected
two-time correlation functions
\begin{equation} 
  C(t,\tw)=\frac{1}{N} \left[ \langle O_\mathrm{s}(t)  
  O_\mathrm{s}(\tw) \rangle \right]
  \quad \mbox{and} \quad  
  C(t,\tw) = \frac{1}{N} \big[ \langle O_\mathrm{d}(t)  
  O_\mathrm{d}(\tw) \rangle - \langle O_\mathrm{d}(t) \rangle  
  \langle O_\mathrm{d}(\tw) \rangle \big] ,
  \label{equ:corO} 
\end{equation} 
for spins and defects, respectively, and the responses 
\begin{equation} 
  R(t,\tw) = \frac{T}{N} \left. \left[ \frac{\delta \langle  
  O_\mathrm{s}(t) \rangle }{\delta h_\mathrm{s}(\tw)} \right] 
  \right|_{h_\mathrm{s}=0} 
  \quad \mbox{and} \quad  
  R(t,\tw)= \frac{T}{N} \left. \left[  \frac{\delta \langle  
  O_\mathrm{d}(t) \rangle}{\delta h_\mathrm{d}(\tw)}  \right]
  \right|_{h_\mathrm{d}=0} ,
  \label{equ:resO} 
\end{equation} 
where $h_\mathrm{s}$ and $h_\mathrm{d}$ are thermodynamically
conjugate to $O_\mathrm{s}$ and $O_\mathrm{d}$, respectively. All
functions are scaled by $N$ to get quantities of order unity. It is
easy to show that, in the thermodynamic limit
$N\to\infty$, Eqs.~(\ref{equ:corO}), (\ref{equ:resO})
become~\cite{peterpeter2}
\begin{equation} 
  C(t,\tw)=\sum_n q_n \cors[n](t,\tw)  
  \quad \mbox{and} \quad  
  R(t,\tw)=\sum_n q_n \ress[n](t,\tw). 
  \label{equ:corchiO} 
\end{equation} 
Here we have used translational invariance (which holds for our quench
from an equilibrium state) to define the distance-dependent
correlation functions
\begin{equation} 
  \begin{array}{rcl} 
    \cors[j-i](t,\tw) & = & \langle s_i(t) s_{j}(\tw) \rangle \quad 
    \mbox{(spins),} \\[1ex] 
    \cord[j-i](t,\tw) & = & \langle s_i(t) s_{i+1}(t) s_{j}(\tw) 
	s_{j+1}(\tw)  
    \rangle - \langle s_i(t) s_{i+1}(t) \rangle \langle s_j(\tw) 
    s_{j+1}(\tw) \rangle \quad \mbox{(defects),} 
  \end{array} 
  \label{equ:corss} 
\end{equation} 
and associated responses 
\begin{equation} 
  \ress[j-i](t,\tw)=T \left. \frac{\delta \langle s_i(t) \rangle} 
  {\delta h_{j}(\tw)}  
  \right|_{h_{j}=0} 
\mbox{(spins)}  \quad \mbox{and} \quad  
  \resd[j-i](t,\tw) = T \left. \frac{\delta \langle s_i(t) s_{i+1}(t) \rangle} 
  {\delta h_{j,j+1}(\tw)}  
  \right|_{h_{j,j+1}=0} \mbox{(defects).} 
  \label{equ:resss} 
\end{equation} 
As usual, $h_j$ and $h_{j,j+1}$ are conjugate to $s_j$ and $s_j
s_{j+1}$, respectively.  Translation invariance also shows that in the
thermodynamic limit $N\to\infty$, the expressions (\ref{equ:corO}),
(\ref{equ:resO}) are self-averaging, i.e.\ independent of the
particular realization of the disorder variables $\epsilon_i$.
 
Analysis of the non-equilibrium FDR for the observables
$O_\mathrm{s}$, $O_\mathrm{d}$ thus requires knowledge of all spin and
defect correlation and response functions (\ref{equ:corss}),
(\ref{equ:resss}). We have tackled this problem in~\cite{peterpeter2}
where we give closed, exact solutions for generic two-time multi-spin
correlation and response functions in the Glauber-Ising chain after a
quench from an arbitrary equilibrium state at temperature
$T_\mathrm{i} > 0$ to any $T \geq 0$. The approach is based on the
closed hierarchy of differential equations for the spin-correlations
$\langle s_{i_1}(t) s_{i_2}(t) \cdots s_{i_k}(t) \rangle$, which we
solved for arbitrary initial correlations $\langle s_{i_1}(0) \cdots
s_{i_k}(0) \rangle$. The key result reads~\cite{peterpeter2}
\begin{eqnarray} 
  & {\displaystyle \langle s_{i_1}(t) \cdots s_{i_k}(t) \rangle =  
    \sum\limits_{l=0}^{\lfloor \frac{k}{2} \rfloor}  
    \sum\limits_{\pi \in  
    \mathcal{P}(l,k)} (-1)^{\pi} \prod\limits_{\lambda=1}^{l}  
    \Hn_{i_{\pi 
    (2\lambda)}-i_{\pi(2\lambda-1)}}(2t) \; \Phi^{(k-2l)}_{(i_{\pi(2l+1)}, 
    \ldots i_{\pi(k)})}(t)} & 
  \label{equ:general} \\ 
  & {\displaystyle \Phi^{(k)}_{\boldsymbol{i}}(t)=\sum\limits_{j_1<\ldots   
    j_k} \Bigg( \sum\limits_{\pi  
    \in \mathcal{S}(k)} (-1)^{\pi} \prod\limits_{\lambda=1}^{k} \me^{-t}  
    \In_{i_\lambda-j_{\pi(\lambda)}} (\gamma t) \Bigg)  
    \langle s_{j_1}(0) \cdots s_{j_k}(0) \rangle } & \label{equ:homogeneous}  
\end{eqnarray} 
Here $\pi$ denotes permutations and $(-1)^\pi$ their sign;
$\mathcal{S}(k)$ is the set of all permutations of $\{1,2,\ldots k\}$
while $\mathcal{P}(l,k)$ is the set of permutations corresponding to
choosing $l$ ordered pairs from the numbers $1,2,\ldots k$ and keeping
the remaining $k-2l$ numbers in ascending order. Explicit expressions
for the functions $\In_n(x)$ and $\Hn_n(x)$ are given
in~\cite{peterpeter2} (for $N \to \infty$ the $\In_n(x)$ are 
modified Bessel functions, see Appendix A). We
also show in~\cite{peterpeter2} that the
evolution of two-time, multi-spin correlation and response
functions is governed by an identical hierarchy of differential
equations, so that these quantities can be obtained from
(\ref{equ:general}), (\ref{equ:homogeneous}) if we substitute the
corresponding equal-time initial conditions in
(\ref{equ:homogeneous}). The latter are just equal-time correlations
---or can be expressed in terms of these for the response functions---
which we know already. For a quench from an equilibrium state this leads 
to explicit results for the two-time multi-spin functions.
As simple 
examples we state in \cite{peterpeter2} the spin and defect functions 
(\ref{equ:corss}), (\ref{equ:resss}) for the quench from 
$T_\mathrm{i} = \infty$ to $T \rightarrow 0$ considered here.
For spins one finds
\begin{eqnarray} 
  \cors[n](t,\tw) &=& \me^{-(t+\tw)} \, \Big\{ \In_n(t+\tw) +  
    \int\limits_0^{2\tw} \upd\tau \;\; \In_n(t+\tw-\tau) \; 
    [\In_0+\In_1](\tau) \Big\}, \label{equ:corsn}\\ 
  \chis[n](t,\tw) &=& \frac{1}{2}\;\me^{-t} \int\limits_{\tw}^t  
    \upd\tau \;\; \me^{-\tau} \, \In_n(t-\tau) \; \left[\In_0 + 
    2 \In_1 + \In_2 \right] (2\tau), \label{equ:chisn} 
\end{eqnarray} 
and for defects 
\begin{eqnarray} 
  \cord[n](t,\tw) &=& \frac{1}{2} \, \me^{-(t+\tw)} \,  
    [\In_{n-1}-\In_{n+1}](t+\tw) \; \int\limits_{t-\tw}^{t+\tw} \upd\tau \;  
    \me^{-\tau} \, [\In_{n-1} - \In_{n+1}](\tau)
\label{equ:cordn}\\ 
  & + & \me^{-2 t} \, \Big\{ \In_n(t-\tw) \; [\In_{n-1}+2 \In_n +  
    \In_{n+1}](t+\tw)  - \me^{-2\tw} \, [(\In_{n-1}+\In_n)  
    (\In_n + \In_{n+1})](t+\tw) \Big\}, \nonumber \\[2ex] 
  \chid[n](t,\tw) &=& \me^{-2t} \, \Big\{ 2 \delta_{n,0} \;  
    [\In_0+\In_1](2t) - \In_n(t-\tw) \; [\In_{n-1}+2\In_{n}+\In_{n+1}](t+\tw)  
    \Big\}.
\label{equ:chidn} 
\end{eqnarray} 
Here and below the short-hand $[\,\ldots\,](x)$ is used to indicate that 
all functions enclosed in the square brackets have the same argument $x$;
$\delta_{n,0}$ is the standard Kronecker delta.  The expressions 
(\ref{equ:chisn}), (\ref{equ:chidn}) 
for the susceptibilities are more convenient than those for the
responses $R_n(t,\tw)=-(\partial/\partial \tw) \chi_n(t,\tw)$ and so
we mostly base the following discussion on them.  We note that while
equations (\ref{equ:corsn}), (\ref{equ:chisn}) have already been
given in various forms, e.g.~\cite{golu1}, we are not aware of any
equivalent in the literature of (\ref{equ:cordn}), (\ref{equ:chidn}).
 
Eqs.~(\ref{equ:corsn})--(\ref{equ:chidn}) will form the basis for our
analysis of the FDR in the $1d$ Ising chain in Secs.~\ref{2spin1d}
and~\ref{4spin1d}.  For now we return to the observables
$O_\mathrm{s}$, $O_\mathrm{d}$ and in particular the choice of the
field covariances $q_n$.  According to~(\ref{equ:corchiO}) we obtain
spin and defect autocorrelation and response functions by choosing
uncorrelated random fields $\epsilon_i$, i.e.\ $q_n = \delta_{n,0}$.
We abbreviate the notation in this case to that used in the
introduction and write $\cors(t,\tw)$ for spin and $\cord(t,\tw)$ for
defect autocorrelations and similarly $\chis(t,\tw)$, $\chid(t,\tw)$
for susceptibilities. Uniform covariances $q_n=1$, on the other hand,
yield full summations over all cross-correlation and response
functions in (\ref{equ:corchiO}).  So $O_\mathrm{s}$ and
$O_\mathrm{d}$ produce just the magnetization and energy,
respectively; we thus use the obvious short-hands $\corm(t,\tw)$,
$\chim(t,\tw)$, and $\core(t,\tw)$, $\chie(t,\tw)$ for this case. It
will turn out that the local ($q_n=\delta_{n,0}$) and global ($q_n=1$)
FDT relations for spin and defect observables are very different.
Therefore we also investigate intermediate choices of $q_n$ that
interpolate between these two extremes.  Two classes of covariances
can be distinguished.  We may interpolate between $q_n=\delta_{n,0}$
and $q_n=1$ by a family of covariances that satisfies $\sum_n |q_n| <
\infty$ for any non-uniform choice of $q_n$; we call the corresponding
fields $\epsilon_i$ short-range correlated. Alternatively, we
can interpolate such that $\sum_n |q_n| = \infty$ as long as the
fields are not completely uncorrelated; we refer to such fields as
infinite-range correlated. In either case the analysis of the
FDR for the correlation and response functions of the associated
observable requires us to evaluate the infinite sums in
(\ref{equ:corchiO}). This can be done conveniently in terms of the
Fourier transforms $q(k)= \mathcal{F}\{q_n\}$,
$C(k;t,\tw)=\mathcal{F}\{ C_n(t,\tw) \}$ and
$\chi(k;t,\tw)=\mathcal{F}\{\chi_n(t,\tw)\}$ where
\begin{equation} 
  \mathcal{F}\{f_n\} = \sum_n f_n \me^{-i n k}  
  \quad \mbox{and} \quad 
  \mathcal{F}^{-1}\{f(k)\} = \int\limits_{-\pi}^{\pi} \frac{\upd k}{2\pi}  
    f(k) \me^{i n k} .
  \label{equ:fourier} 
\end{equation} 
In Appendix~\ref{sec:fourier} we state the Fourier transforms of  
(\ref{equ:corsn})--(\ref{equ:chidn}), in terms of which 
(\ref{equ:corchiO}) becomes 
\begin{equation}  
  C(t,\tw)=\int\limits_{-\pi}^{\pi} \frac{\upd k}{2\pi} \; 
  q(k) C(k;t,\tw)  
  \quad \mbox{and} \quad  
  \chi(t,\tw)=\int\limits_{-\pi}^{\pi} \frac{\upd k}{2\pi}  
  \; q(k) \chi(k;t,\tw) .
  \label{equ:corchiOfourier} 
\end{equation} 
An explicit example of a family of short-range correlated fields,
parameterized by $a>0$, is given by the Lorentzian covariances
\begin{equation} 
  q_{\lorenz,n}=\frac{a^2}{a^2+n^2} \quad \Leftrightarrow \quad  
  q_{\lorenz}(k)=\frac{a \pi}{\sinh a \pi} \cosh a (\pi-|k|) .
  \label{equ:lorenzQ} 
\end{equation} 
The transform (\ref{equ:lorenzQ}) can be found  
in any table of Fourier transforms. Eq.~(\ref{equ:lorenzQ}) indeed
defines short-range correlated fields: since $q_{\lorenz,n} > 0$ the
criterion becomes 
$N_\mathrm{c}\equiv\sum_n q_{\lorenz,n} < \infty$ which is satisfied since 
$N_\mathrm{c} = q_{\lorenz}(0) = a \pi \coth a \pi$. 
By varying $a$ we can also smoothly tune our observables  
between local ($q_{\lorenz,n} \rightarrow \delta_{n,0}$ as  
$a \rightarrow 0$) and global ($q_{\lorenz,n} \rightarrow 1$  
for $a \rightarrow \infty$) ones. We denote the corresponding  
correlations and susceptibilities by 
$\cors[\lorenz](t,\tw)$ and $\chis[\lorenz](t,\tw)$. 
An example of covariances that yield infinite-range correlated fields  
is 
\begin{equation} 
  q_{\power,n}=(-1)^n \frac{\Gamma^2(\frac{1+\alpha}{2})} 
  {\Gamma(\frac{1+\alpha}{2}-n)  
  \Gamma(\frac{1+\alpha}{2}+n)} 
  \quad \Leftrightarrow \quad 
  q_{\power}(k)=\frac{\Gamma^2(\frac{1+\alpha}{2})}{2^{1-\alpha}  
  \Gamma(\alpha)} \, \left| \sin {\textstyle \frac{k}{2}} \right|^{\alpha-1} ,
  \label{equ:powerQ} 
\end{equation} 
where $0<\alpha<1$ and $\Gamma(x)$ is the Gamma
function~\cite{mathbook}. It is clear from (\ref{equ:powerQ}) that
$q_{\power,n}$ is even in $n$ and $q_{\power,0}=1$. We show in
Appendix~\ref{sec:powerQ} that for $\alpha \to 1$ we get
$q_{\power,n}=\delta_{n,0}$ while $\alpha \to 0$ gives
$q_{\power,n}=1$. We also prove there that $q_{\power,n}$ decreases
monotonically as $|n|$ increases, decaying asymptotically as a
power-law $q_{\power,n} \sim |n|^{-\alpha}$, and that indeed
$\mathcal{F}^{-1}\{ q_{\power}(k) \} = q_{\power,n}$. The reverse
transform $\mathcal{F}\{ q_{\power,n} \}$ does not converge in the
usual sense, but this is not necessary for equivalence of
(\ref{equ:corchiO}) and (\ref{equ:corchiOfourier}).  So
(\ref{equ:powerQ}) again allows us to interpolate smoothly between
local and global observables, but in such a way that $\sum_n
q_{\power,n} = \infty$ for any $\alpha \in [0,1[$.  The correlations
and susceptibilities for the observables defined by fields
$\epsilon_i$ with the power-law covariances (\ref{equ:powerQ}) are
denoted by $\cors[\power](t,\tw)$ and $\chis[\power](t,\tw)$ below.

\subsection{Spin observables} 
\label{2spin1d} 
 
\subsubsection{Random field: Incoherent functions} 
\label{2spin1ds} 
 
The FDT violation for the spin autocorrelation and response function
has already been studied in detail in \cite{golu1,zaza}. In particular
it was shown that the FD plot approaches a non-trivial limit curve in
the aging regime, with $\Xs^\infty=\frac{1}{2}$.  We can easily
recover the existing results for $\cors(t,\tw)$, $\chis(t,\tw)$ from
our exact solutions (\ref{equ:corsn}), (\ref{equ:chisn}) by setting
$n=0$. It is useful to focus on the aging limit. Formally this is an
asymptotic expansion in the limit $t,\tw \to \infty$ with
$\epsilon \leq \tw/t \leq 1-\delta$ fixed and $\epsilon,\delta > 0$, to
ensure that $t,\tw$  and $\dt$ all diverge and are of the
same order. In this limit the asymptotic expansion (\ref{equ:Inasym})
for modified Bessel functions yields immediately
\begin{eqnarray} 
\cors(t,\tw) & \sim & \frac{2}{\pi} \arcsin\sqrt{\frac{2\tw}{t+\tw}},  
  \label{equ:corsasymp} \\ 
\chis(t,\tw) & \sim & \frac{\sqrt{2}}{\pi} \arccos \sqrt{\frac{\tw}{t}}. 
  \label{equ:chisasymp} 
\end{eqnarray} 
Here and below, the `$\sim$' sign denotes results which are
asymptotically exact in the aging limit.  The limit FD-plot
corresponding to (\ref{equ:corsasymp}), (\ref{equ:chisasymp}) is
contained in Fig.~\ref{fig:spinpowerlimit} below and the associated
FDR is a function of the time ratio $\tw/t$ only,
\begin{equation} 
  \Xs(t,\tw) \sim \frac{t+\tw}{2t} .
  \label{equ:Xsasymp} 
\end{equation} 
It shows a continuous crossover from $\Xs(t,\tw) = 1$ for $\dt \ll
\tw$ to $\Xs(t,\tw) = \Xs^\infty=\frac{1}{2}$ for $\dt \gg \tw$. We
note that the aging expansion of the spin correlations and
susceptibilities (\ref{equ:corsn}), (\ref{equ:chisn}) is dominated by
the leading term of the asymptotic series (\ref{equ:Inasym}) for the
modified Bessel functions, which is independent of the order $n$.
Therefore (\ref{equ:corsasymp}), (\ref{equ:chisasymp}) in fact apply
to all finite-distance spin cross-correlations and susceptibilities
$\cors[n](t,\tw)$, $\chis[n](t,\tw)$. Consequently, the latter produce
the same limiting FD plot and FDR (\ref{equ:Xsasymp}) as for $n=0$.

\subsubsection{Uniform field: Coherent functions} 
\label{2spin1dm} 
 
As described, the uniform field effectively allows us to study the FDR
for the magnetization. The corresponding correlation and
susceptibility are most conveniently obtained from the Fourier
transforms (\ref{equ:corsfourier}), (\ref{equ:chisfourier}) by setting
$k=0$; the time integrals appearing in $C(k;t,\tw)$, $\chi(k;t,\tw)$
can then be solved. One finds
\begin{eqnarray} 
  \corm(t,\tw) & = & \me^{-2 \tw} \left\{\In_0(2\tw)+4 \tw [\In_0 + 
    \In_1](2 \tw) \right\} , \label{equ:corm} \\ 
  \resm(t,\tw) & = & \frac{1}{2} \me^{-2\tw} [\In_0+2\In_1+\In_2](2\tw) .
    \label{equ:resm}  
\end{eqnarray} 
We have given the response $R=-\partial \chi/\partial \tw$ here rather
than the susceptibility because it has a simpler form. Note that
both the correlation and response function (\ref{equ:corm}),
(\ref{equ:resm}) are independent of $t$. This can be understood from
the fact that for $T=0$ the magnetization $m=(1/N)\sum_i s_i$ performs
a random walk with step size $\pm 2/N$ and a time-dependent rate
$\frac{1}{2}Nc$, where $c$ is the concentration of domain walls.  The
latter can be obtained explicitly from (\ref{equ:corsn}) by setting
$t=\tw$ and $n=1$, since $C_1(\tw,\tw) = \langle s_i(\tw)
s_{i+1}(\tw) \rangle = 1-2c(\tw)$. This gives~\cite{defectformula}
\begin{equation} 
  c(\tw)=\frac{1}{2} \me^{-2\tw} [\In_0+\In_1](2\tw) .
  \label{equ:defectdens} 
\end{equation} 
Now, since the random walk of $m$ is unbiased, $\corm(t,\tw)=N\langle
m(t)m(\tw)\rangle=\corm(\tw,\tw)$ follows immediately. Also,
$\corm(\tw,\tw)$ will grow with rate $2ND(\tw)$, where
$D=(2/N)^2\frac{1}{2}Nc=2c/N$ is the diffusion constant. One should
thus have $\partial \corm(\tw,\tw)/\partial\tw=4c(\tw)$, and from
(\ref{equ:corm}), (\ref{equ:defectdens}) one verifies that this is
indeed the case.  Similar arguments apply to the response
$\resm(t,\tw)$.  Brief application of a field at $\tw$ biases the
domain wall motion and hence the random walk of $m$; thereafter the
random walk is again unbiased and so the response must be
$t$-independent.  The momentary bias in the random walk rates contains
the domain wall concentration $c(\tw)$ as an overall factor, and
consistent with this expectation one gets $\resm(t,\tw) \sim 2 c(\tw)$
asymptotically.

Since both $\corm(t,\tw)$ and $\resm(t,\tw)$  
are functions of $\tw$ only, this also applies to the FDR 
\begin{equation} 
  \Xm(\tw)=\frac{[\In_0+2\In_1+\In_2](2\tw)}{4[\In_0+ 
    \In_1](2\tw)}, \label{equ:Xm} 
\end{equation} 
which crosses over from the initial value $\Xm(0)=\frac{1}{4}$ to
$\frac{1}{2}$ on an $\mathcal{O}(1)$ time scale. So, apart from a
transient after the quench, we measure $\Xm(\tw) = \frac{1}{2}$ for
all $t \geq \tw \gg 1$; in particular the limiting value $\Xm(\tw)
\sim \Xm^\infty = \Xs^\infty = 1/2$ in the aging regime is identical
to that for the incoherent spin observables. Note that there is no
quasi-equilibrium regime with $\Xm=1$ for $\dt \ll \tw$. The
corresponding FD plot converges to a straight line of slope
$\frac{1}{2}$ (see Fig.~\ref{fig:spinpowerlimit} below).

\subsubsection{Short-range correlated field} 
\label{2spin1dL} 
 
Next we investigate the effect of short-range correlations in the
random fields $\epsilon_i$ on the FDR. The correlations and
susceptibilities of the corresponding observables may be obtained
either from a real-space summation (\ref{equ:corchiO}) or an
integration in the Fourier representation
(\ref{equ:corchiOfourier}). Using the latter, we note first that the
short-range criterion $\sum_n |q_n| < \infty$ for the covariances
implies that $q(k) = \mathcal{F}\{ q_n \} = \sum_n q_n \me^{-i n k}$
is a continuous function. The Fourier transforms $C(k;t,\tw)$,
$\chi(k;t,\tw)$, on the other hand, satisfy
\begin{equation} 
  \frac{C(k;t,\tw)}{\cors(t,\tw)} \rightarrow 2\pi\tilde{\delta}(k) \quad  
  \mbox{and} \quad  
  \frac{\chi(k;t,\tw)}{\chis(t,\tw)} \rightarrow 2\pi\tilde{\delta}(k) 
  \label{equ:spincorchidelta} 
\end{equation} 
in the aging limit, where $\tilde\delta(\cdot)$ is a $2\pi$-periodic 
version of the ordinary Dirac delta. The normalizations of the 
right-hand sides of (\ref{equ:spincorchidelta}) are clear since 
$\cors(t,\tw)$, for  
instance, is given by the Fourier integral (\ref{equ:fourier}) over  
$C(k;t,\tw)$ for $n=0$. And since $\cors(t,\tw)$ and $\chis(t,\tw)$  
are $\mathcal{O}(1)$ functions of 
$\tw/t$ in the aging limit, while $C(k;t,\tw)$, $\chi(k;t,\tw)$ 
vanish in the same limit for any $k$  
that is not a multiple of $2\pi$, Eq.~(\ref{equ:spincorchidelta}) 
follows. This in turn implies that for any short-range   
correlated field 
\begin{equation} 
C(t,\tw) \sim  N_\mathrm{c} \cors(t,\tw)  
  \quad \mbox{and} \quad  
\chi(t,\tw) \sim N_\mathrm{c} \chis(t,\tw)  
  \label{equ:shortfieldasym} 
\end{equation} 
provided that $N_\mathrm{c}=q(0)=\sum_n q_n$, which estimates the number of 
lattice sizes over which field correlations extend, is nonzero. So in 
the aging limit the 
correlations and susceptibilities are ultimately just  
proportional to (\ref{equ:corsasymp}), (\ref{equ:chisasymp}) and hence  
yield the same FDR and FD plot as the local spin functions. This statement  
can equivalently be made in real space, based on convergence of the  
series $q_n$ and the fact that all finite-distance cross-correlation  
and response functions behave asymptotically as (\ref{equ:corsasymp}),  
(\ref{equ:chisasymp}).  
 
\begin{figure}[htb]
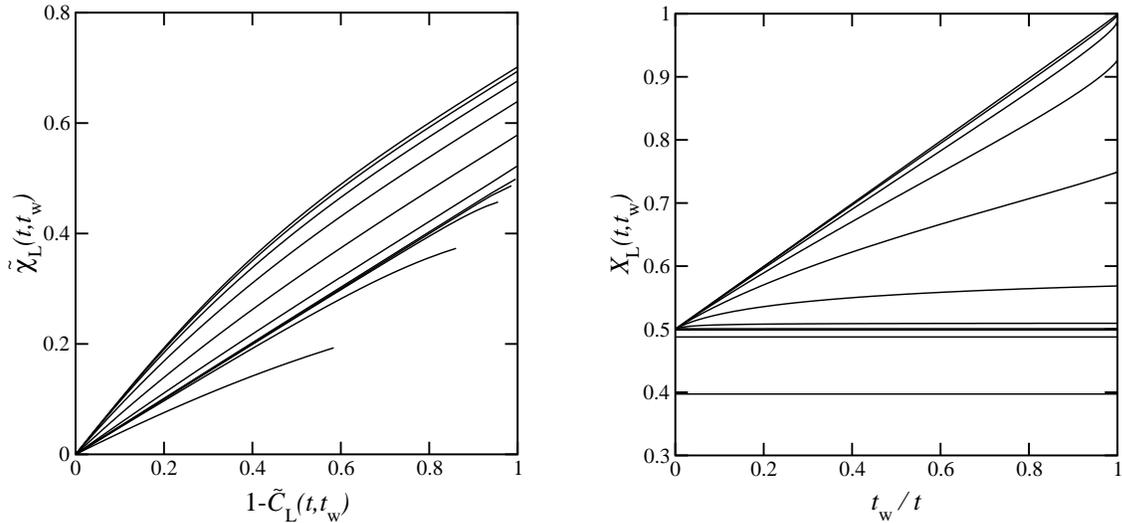
 
  \begin{picture}(15,7) 
    \put(0,0){\epsfig{file=Lorenz1000FDT.eps,scale=0.4,clip}} 
    \put(8,0){\epsfig{file=Lorenz1000X.eps  ,scale=0.4,clip}} 
  \end{picture} 
  \caption{\label{fig:spinlorenz} Normalized FD plot (left) and the  
    corresponding FDR versus $\tw/t$ (right) for a random field with  
    Lorentzian covariances (\ref{equ:lorenzQ}) with $a=10^3$, 
    correlated over $N_c=a \pi \coth a \pi\approx a\pi$ sites. In the  
    FD plot, $t$ is fixed for each curve and varies over the range 
    $10^0,10^1,\dots10^{10}$ 
    (bottom to top). The 
    lines first converge towards the straight line with slope $\frac{1}{2}$  
    corresponding to a coherent observable (the magnetization) but 
    eventually, for $t\geq 10^5$, 
    cross over to the limit plot for uncorrelated fields. This behaviour is  
    also reflected in  
    the evolution of $\Xs[\lorenz](t,\tw)$. There, however, we have  
    the freedom to fix either $t$ or $\tw$. The plot shows the  
    case of fixed $\tw$, which is more convenient for comparison 
    with simulations, for $\tw=10^0,\ldots 10^{10}$ (bottom to top).} 
\end{figure}

At finite times, however, we find a crossover between two dynamical  
regimes. A scaling analysis shows that the peaks in $C(k;t,\tw)$ and  
$\chi(k;t,\tw)$ at $k=0$ have widths $t^{-1/2}$ and $\dt^{-1/2}$,  
respectively. Correspondingly, we have growing length  
scales in real space. These are $\ell_C \approx t^{1/2}$ for correlations,  
corresponding to the typical domain size, but $\ell_\chi  
\approx \dt^{1/2}$ for the response which reflects the fact that  
perturbations spread diffusively. When $\ell_C, \ell_\chi \gg N_{\mathrm{c}}$, 
one has an effectively local observable 
and we are in the asymptotic regime (\ref{equ:shortfieldasym}). If, however,  
$\ell_C, \ell_\chi \ll N_{\mathrm{c}}$, the fields $\epsilon_i$ are 
correlated over distances much longer than the dynamical length 
scales, giving an effectively uniform field. One thus expects to get 
an FD plot similar to that obtained for  
the magnetization. The illustration of the crossover in 
Fig.~\ref{fig:spinlorenz}, obtained by numerical integration of 
(\ref{equ:corchiOfourier}), shows that this is indeed the case. 

\subsubsection{Infinite-range correlated field} 
\label{2spin1dP} 
 
\begin{figure}[htb]
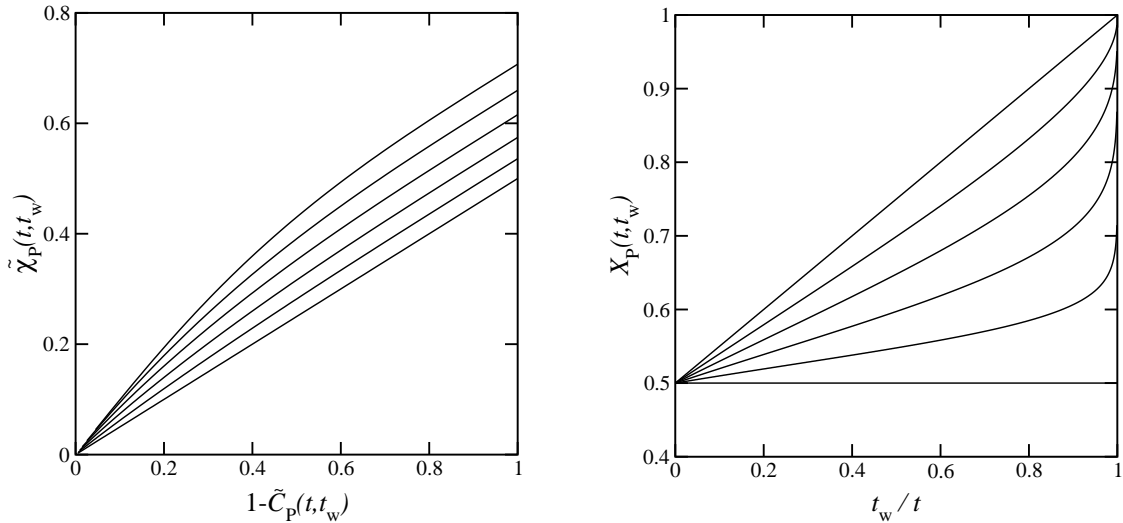
 
  \begin{picture}(15,7) 
    \put(0,0){\epsfig{file=PowerLimitFDT.eps,scale=0.4,clip}} 
    \put(8,0){\epsfig{file=PowerLimitX.eps  ,scale=0.4,clip}} 
  \end{picture} 
  \caption{\label{fig:spinpowerlimit} Normalized FD-plots (left)  
    and the corresponding FDR versus $\tw/t$ (right) in the aging  
    limit $t,\tw \to \infty$. The bottom curves  
    in both plots are for the magnetization (Sec.~\protect\ref{2spin1dm}) and 
    coincide with the uniform field limit $\alpha \to 0$ of power-law field 
    covariances. The intermediate curves are for power-laws  
    (Sec.~\protect\ref{2spin1dP}) with exponents $\alpha = 0.2, 0.4, 0.6, 0.8$  
    (bottom to top). The top curves represent the random field limit  
    $\alpha \to 1$ for power-laws, and apply also to any short-range 
    correlated field (Sec.~\protect\ref{2spin1dL}) with $N_\mathrm{c} 
    \neq 0$ or, in the extreme case, the incoherent   
    functions of Sec.~\protect\ref{2spin1ds}.} 
\end{figure}

For infinite-range correlated fields one cannot use simple scaling
arguments since the correlations and susceptibilities contain
contributions from all length scales. For the power-law covariances
(\ref{equ:powerQ}) introduced above, this is reflected in the
singularity of $q_\power(k)$ at $k=0$. Therefore we have to analyze
the full expressions for $\cors[\power](t,\tw)$ and
$\chis[\power](t,\tw)$ that follow from (\ref{equ:corchiOfourier})
after substitution of $q_{\power}(k)$, $C(k;t,\tw)$ and
$\chi(k;t,\tw)$. Fortunately, for the particular choice of
$q_{\power}(k)$ the results may be expressed in terms of single
integrals of the form
\begin{eqnarray} 
  \cors[\power](t,\tw) & \!\! = & \!\! \me^{-2(t+\tw)} \Bigg\{ \! 
    {_1}\mathrm{F}{_1}\left( {\textstyle \frac{1}{2},\frac{1+\alpha}{2}; 
    2(t\!+\!\tw) } \right) \! + \!\!\! 
    \int\limits_0^{2\tw} \!\! \upd\tau \, \me^\tau  
    {_1}\mathrm{F}{_1}  
    \left( {\textstyle \frac{1}{2}, 
    \frac{1+\alpha}{2};2(t\!+\!\tw\!-\!\tau) } \right) [\In_0+\In_1](\tau) \! 
    \Bigg\}, \label{equ:corp} \\ 
  \chis[\power](t,\tw) & \!\! = & \!\! \frac{1}{2} \, \me^{-2t} \!\! 
    \int\limits_{\tw}^t \!\! 
    \upd\tau \,  
    {_1}\mathrm{F}{_1} 
    \left( {\textstyle \frac{1}{2},\frac{1+\alpha}{2};2(t\!-\!\tau)} \right)  
    [\In_0+2\In_1+\In_2](2\tau), \label{equ:chip} 
\end{eqnarray} 
where ${_1}\mathrm{F}{_1}(\alpha,\gamma;z)$ is the confluent
hypergeometric function \cite{mathbook}. Equations.~(\ref{equ:corp}),
(\ref{equ:chip}) are exact and can be used to study the FD plots and
the FDR numerically. However, in the aging limit asymptotic expansions
may be substituted for the non-elementary functions and significant
simplifications are possible. One finds
\begin{eqnarray} 
  \cors[\power](t,\tw) & \sim & \frac{2^{\frac{1-\alpha}{2}}}{\pi} \,  
    \Gamma\left( {\textstyle \frac{1+\alpha}{2} } \right) \,  
    (t+\tw)^{\frac{1-\alpha}{2}} \mathrm{B} 
    ({\textstyle \frac{1}{2}, 1-\frac{\alpha}{2}; \frac{2\tw}{t+\tw} }) ,
    \label{equ:corspasym} \\ 
  \chis[\power](t,\tw) & \sim & \frac{2^{-\frac{\alpha}{2}}}{\pi} \,  
    \Gamma\left({\textstyle \frac{1+\alpha}{2} } \right) \,  
    t^{\frac{1-\alpha}{2}} 
    \left[ \mathrm{B}({\textstyle \frac{1}{2}, 1-\frac{\alpha}{2} }) -  
    \mathrm{B}({\textstyle \frac{1}{2}, 1-\frac{\alpha}{2}; \frac{\tw}{t} }) 
    \right] ,
    \label{equ:chispasym} 
\end{eqnarray} 
where $\mathrm{B}(p,q;x)$ is the incomplete Beta function
$\mathrm{B}(p,q;x) = \int_0^x \upd u \, u^{p-1} (1-u)^{q-1}$ and
$\mathrm{B}(p,q)=\mathrm{B}(p,q;1)$ is the complete
one~\cite{mathbook}.  In the random field limit, $\alpha \to 1$, we
recover the expansions (\ref{equ:corsasymp}), (\ref{equ:chisasymp})
for the incoherent functions since
$\mathrm{B}(\frac{1}{2},\frac{1}{2};x)=2 \arcsin \sqrt{x}$, whereas
the uniform field limit $\alpha \to 0$ can be shown to coincide, using
$\mathrm{B}(\frac{1}{2},1;x) = 2 \sqrt{x}$, with the asymptotic
expansions of (\ref{equ:corm}), (\ref{equ:resm}) for the coherent
functions.  So the power-law covariances (\ref{equ:powerQ}) indeed
allow us to interpolate between the coherent and incoherent
observables. For intermediate exponents $0<\alpha<1$ the fluctuations
in the observable $O_\mathrm{s}$ grow as $\tw^{(1-\alpha)/2}$ and the
two-time correlation (\ref{equ:corspasym}) has a plateau at a
corresponding value for $\dt \ll \tw$; for $\dt \gg \tw$ it decays as
$\tw^{(1-\alpha)/2} (\tw/\dt)^{\alpha/2}$. For the susceptibility we
deduce from (\ref{equ:chispasym}) a $\dt^{(1-\alpha)/2}
(\dt/\tw)^{1/2}$ growth for $\dt \ll \tw$ that crosses over to
$\dt^{(1-\alpha)/2}$ for $\dt \gg \tw$. Fig.~\ref{fig:spinpowerlimit}
shows exact FD limit plots that follow from (\ref{equ:corspasym}),
(\ref{equ:chispasym}). The associated FDR may be obtained from
(\ref{equ:corspasym}), (\ref{equ:chispasym}) as
\begin{equation} 
  \Xs[\power](t,\tw) \sim \Bigg\{ \frac{2t}{t+\tw}+\frac{1-\alpha}{2} 
    \sqrt{\frac{2\tw}{t+\tw}}  
    \left(\frac{t-\tw}{t+\tw}\right)^{\frac{\alpha}{2}}  
    \mathrm{B} 
    ({\textstyle \frac{1}{2}, 1-\frac{\alpha}{2}; \frac{2\tw}{t+\tw} }) 
    \Bigg\}^{-1}. \label{equ:Xspasym} 
\end{equation} 
In principle one should first differentiate (\ref{equ:corp}),
(\ref{equ:chip}) to obtain $\ress[\power](t,\tw)$ and
$({\partial}/{\partial \tw})\cors[\power](t,\tw)$ and then
perform the aging expansion, but this turns out to give the same
result. Equation~(\ref{equ:Xspasym}) is a function of $\tw/t$ only and
interpolates between the FDR (\ref{equ:Xsasymp}) for the local spin
observables ($\alpha \to 1$) and the constant $\Xm^\infty =
\frac{1}{2}$ for the magnetization ($\alpha \to 0$). Plots of
$\Xs[\power](t,\tw)$ for various powers $\alpha$ are also shown in
Fig.~\ref{fig:spinpowerlimit}. It is remarkable that the FDR again
crosses over from $\Xs[\power](t,\tw)=1$ for $\dt \ll \tw$ to
$\Xs[\power]^\infty = \frac{1}{2}$ for $\dt \gg \tw$, independently of
the power-law exponent $\alpha$.

\subsubsection{Harmonically correlated fields and $X^\infty$} 
\label{2spin1dX}
 
The explicit examples given in Sec.~\ref{2spin1ds}-\ref{2spin1dP}
suggest that $X^\infty = \frac{1}{2}$ is a generic feature for the
spin-observables $O_\mathrm{s}$ defined in (\ref{equ:observables}).
To show that this is indeed true, we start from the fact that for a
general observable $O_\mathrm{s}$ the correlation and susceptibility
---and hence $(\partial/\partial \tw) C(t,\tw)$ and $R(t,\tw)$--- may
be written in the form (\ref{equ:corchiOfourier}). By introducing a
generalized FDT for the Fourier modes
\begin{equation} 
  R(k;t,\tw) = X(k;t,\tw) \frac{\partial}{\partial \tw} C(k;t,\tw) ,
  \label{equ:FDTsfourier} 
\end{equation} 
we may express $R(k;t,\tw)$ via (\ref{equ:FDTsfourier}) and thereby
obtain the following representation for the FDR $X(t,\tw)$ associated
with a generic spin-observable $O_\mathrm{s}$:
\begin{equation} 
  X(t,\tw)=\frac{\int\limits_{-\pi}^{\pi} \frac{\upd k}{2\pi}  
    \, X(k;t,\tw) \, q(k)  
    \frac{\partial}{\partial \tw} C(k;t,\tw)}{ 
    \int\limits_{-\pi}^{\pi} \frac{\upd k}{2\pi} \, q(k) \frac{\partial} 
    {\partial \tw} C(k;t,\tw)} .
  \label{equ:XsQ} 
\end{equation} 
This means that $X(t,\tw)$ may be considered as the average of
$X(k;t,\tw)$ over the normalized distribution of $q(k)
(\partial/\partial \tw)C(k;t,\tw)$ on $k \in [-\pi,\pi]$. The FDR for
Fourier modes follows from (\ref{equ:FDTsfourier}) and the expressions
(\ref{equ:corsfourier}), (\ref{equ:chisfourier}) for $C(k;t,\tw)$,
$\chi(k;t,\tw)$ as
\begin{equation} 
  X(k;\tw) = \frac{[\In_0+2\In_1+\In_2](2\tw)}{4 [\In_0+\In_1](2\tw) -  
    2 (1-\cos k) \left\{ \me^{2\tw \cos k} + \int_0^{2\tw}  
    \upd\tau \, \me^{(2\tw-\tau) \cos k} [\In_0+\In_1](\tau) \right\}}  ,
  \label{equ:Xsk} 
\end{equation} 
and is a function of $\tw$ and $k$ only. For $k=0$, (\ref{equ:Xsk})
reduces to the FDR for the magnetization $\Xm(\tw)$ (\ref{equ:Xm}) and
hence $X(0;\tw) \approx \frac{1}{2}$ for $\tw \gg 1$. A scaling
analysis of (\ref{equ:Xsk}) shows that for $|k| \ll \pi$ and $\tw \gg
1$ we get $X(k;\tw) \approx X(k^2 \tw)$ with $X(k^2 \tw) \approx
\frac{1}{2}$ for $k^2 \tw \ll 1$ and $X(k^2 \tw) \approx 1$ when $k^2
\tw \gg 1$.  So (\ref{equ:Xsk}) reflects the successive equilibration
of increasing length scales.
 
Now we can return to the FDR (\ref{equ:XsQ}) for the observable
$O_\mathrm{s}$. For the magnetization ---being the coherent observable---
we have $q(k) = 2\pi \tilde{\delta}(k)$ and (\ref{equ:XsQ}) reduces
to the trivial identity $\Xm(\tw) = X(0;\tw)$. In physical terms, by
selecting the coherent observable we only measure the FDR associated
with the infinite length scale. For other spin-observables, being
characterized by the function $q(k)$, the FDR $X(t,\tw)$ contains
contributions from all length scales.  For the long-time limit
$X^\infty$, however, the situation simplifies because
$(\partial/\partial \tw) C(k;t,\tw)$ develops an infinitely sharp peak
at $k=0$ as $t \to \infty$. This can be verified by a scaling analysis
of (\ref{equ:corsfourier}). For sufficiently well-behaved functions
$q(k)$, the normalized version of the distribution $q(k)
(\partial/\partial \tw)C(k;t,\tw)$ thus becomes a realization of
$\tilde{\delta}(k)$ and we get $X(t,\tw) \to X(0;\tw)$ as $t \to
\infty$. Taking the limit $\tw \to \infty$ then shows that $X^\infty =
\frac{1}{2}$, as claimed. So for a generic spin observable, $X^\infty$
again just gives the FDR associated with the infinite length scale.
The only exception occurs when this contribution is explicitly
suppressed. An example of the latter case would be harmonically
correlated fields, $q_n = \cos n p$
with $0<p<\pi$: for such observables $X(t,\tw)=X(p;\tw)$ and hence
$X^\infty = 1$.

\subsection{Defect observables} 
\label{4spin1d} 
 
\subsubsection{Random field: Incoherent functions} 
\label{4spin1dd} 
 
The defect observable $O_\mathrm{d}$ given in (\ref{equ:observables})
with random, uncorrelated fields $\epsilon_i$ allows us to study the
FDT violation for local defect correlations and
susceptibilities. These follow from (\ref{equ:cordn}),
(\ref{equ:chidn}) by setting $n=0$, giving
\begin{eqnarray} 
  \cord(t,\tw) & = & 2\me^{-2 t} \In_0(t-\tw) [\In_0 + \In_1](t+\tw)  
    - \me^{-2 (t+\tw)} [\In_0+\In_1]^2(t+\tw), \label{equ:cord} \\[1ex] 
  \chid(t,\tw) & = & 2 \me^{-2t} \big\{ [\In_0+\In_1](2t)  
    - \In_0(t-\tw) [\In_0+\In_1](t+\tw) \big\} . \label{equ:chid} 
\end{eqnarray} 
These results can be written in a more physically intuitive way in
terms of the concentration of domain walls $c(t)$,
Eq.~(\ref{equ:defectdens}), and the return probability
$p_\mathrm{r}(\tau) = \me^{-\tau} \In_0(\tau)$ of a continuous-time
random walker on a discrete, one-dimensional lattice~\cite{walker}.
Expressing all time dependencies in (\ref{equ:cord}), (\ref{equ:chid})
via $c(t)$ and $p_\mathrm{r}(\tau)$ yields the exact identities
\begin{eqnarray} 
  \cord(t,\tw) & = & 4 c\left( {\textstyle \frac{t+\tw}{2}} \right)  
    [p_\mathrm{r}(t-\tw)-c\left( {\textstyle \frac{t+\tw}{2} } \right)]  ,
    \label{equ:cordrndw} \\ 
  \chid(t,\tw) & = & 4 [ c(t) - p_\mathrm{r}(t-\tw)  
    c\left({\textstyle \frac{t+\tw}{2}}\right)]  .
    \label{equ:chidrndw} 
\end{eqnarray} 
The fact that we find random walk-related quantities does not come as
a surprise given that there is an exact mapping of zero temperature
Glauber dynamics in the Ising chain to a diffusion-limited pair
annihilation (DLPA) process~\cite{MAP}.  The mapping follows by
assigning to each bond $(i,i+1)$ the `particle' occupation number $b_i
= \frac{1}{2}(1-s_i s_{i+1}) \in \{0,1\}$ which signals the presence
or absence of a domain wall. Glauber dynamics for the spins
corresponds to independent random walks for the particles and
coalescence of domains of aligned spins yields particle
pair-annihilation.
 
It follows from the definition (\ref{equ:corss}) of the defect
autocorrelation that $\cord(t,\tw) = 4 \left( \langle b_i(t) b_i(\tw)
\rangle - \langle b_i(t) \rangle \langle b_i(\tw) \rangle \right)$ in
fact also describes the particle autocorrelation in the DLPA
process. We note that (\ref{equ:cordrndw}) is a non-trivial result.
Assuming as in Ref.~\cite{juanpe1} that the autocorrelation of the
fraction $c(t)$ of particles that still exist at time $t$ is given by
$p_\mathrm{r}(t-\tw)$ and that these particles are uncorrelated with
the fraction $c(\tw)-c(t)$ of particles that have disappeared via
annihilation, one would conclude $\langle b_i(t) b_i(\tw) \rangle =
c(t) p_\mathrm{r}(t-\tw) + c(t) (c(\tw)-c(t))$ and hence $\cord(t,\tw)
= 4 c(t) (p_\mathrm{r}(t-\tw) - c(t))$. This obviously differs from
the exact solution (\ref{equ:cordrndw}). As an approximation it holds
for $\dt \ll \tw$, but breaks down for $\dt \gg \tw$ where
(\ref{equ:cordrndw}) yields $\cord(t,\tw) \approx 2\tw/(\pi \dt^2)$
whereas the approximation gives $\cord(t,\tw) \approx (\sqrt{2}-1)/
(\pi \dt)$. This shows that two-time correlations in $\cord(t,\tw)$
build up via a rather subtle mechanism, the explanation of which in
terms of DLPA would probably require knowledge of the inter-particle
(i.e.\ domain size) distribution. Similarly, it appears that the
result (\ref{equ:chidrndw}) cannot be obtained in a straightforward
way.

\begin{figure}[htb]
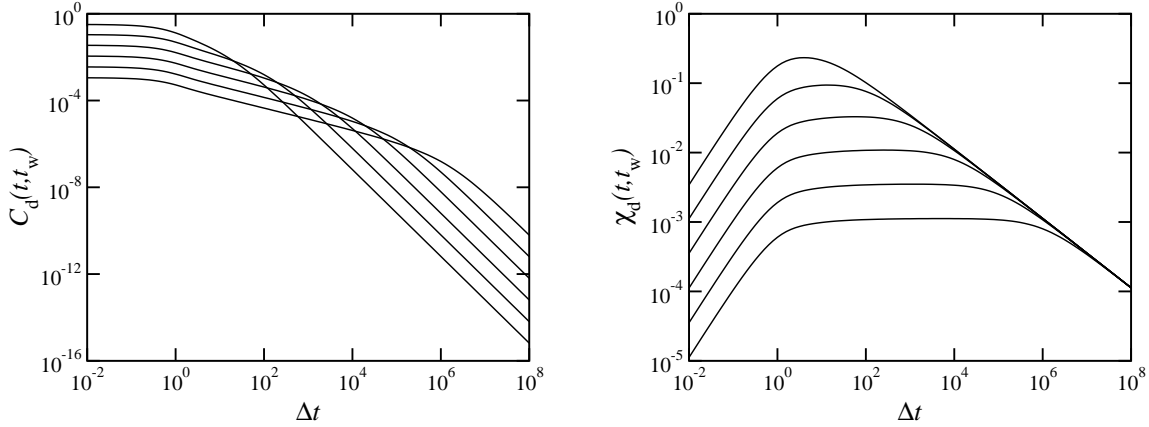
 
  \begin{picture}(15,5.5) 
    \put(0,-0.2){\epsfig{file=DefectCorrelation.eps    ,scale=0.4,clip}} 
    \put(8,-0.2){\epsfig{file=DefectSusceptibility.eps ,scale=0.4,clip}} 
  \end{picture} 
  \caption{\label{fig:defectfunctions} Defect autocorrelation (left) and 
    susceptibility (right) versus $\dt$ for waiting-times $\tw = 10^1, 10^2, 
    \ldots 10^6$. Increasing waiting times correspond to decreasing values 
    in the plot for small $\dt$.}
\end{figure} 
Now we turn to the dynamics of $\cord(t,\tw)$, $\chid(t,\tw)$ --
examples of which are shown in Fig.~\ref{fig:defectfunctions} -- as
given by (\ref{equ:cordrndw}), (\ref{equ:chidrndw}). The equal-time
value of $\cord(\tw,\tw)=4c(\tw)(1-c(\tw)) \approx 4c(\tw)\approx
2/\sqrt{\pi \tw}$ for $\tw \gg 1$ decreases with $\tw$, reflecting the
decreasing number of particles in the DLPA process (or domain walls in
the spin chain). In the regime $\dt \ll \tw$ the two-time correlation
$\cord(t,\tw) \approx 4 c(\tw) p_\mathrm{r}(\dt)$ drops from its
initial value due to the random walk motion of the particles around
their initial positions at $\tw$. In the aging limit of large $\dt$
and $\tw$ one has the expansion $\cord(t,\tw) \sim 2/(\pi
\sqrt{t+\tw}) \left( 1/\sqrt{t-\tw} - 1/\sqrt{t+\tw} \right)$.  This
crosses over from $\cord(t,\tw) \approx 2/(\pi \sqrt{2 \dt \, \tw})$
for $\dt \ll \tw$, where it connects smoothly to the initial drop for
$\dt$ of ${\mathcal{O}}(1)$ since $p_\mathrm{r}(\dt) \approx 1/\sqrt{2
\pi \dt}$ for large $\dt$, to $\cord(t,\tw) \approx 2\tw/(\pi \dt^2)$
for $\dt \gg \tw$. The integrated response $\chid(t,\tw)$ is
non-monotonic in $\dt$ and increases on an $\mathcal{O}(1)$ time scale
in $\dt$ from its initial value $\chid(\tw,\tw)=0$ to a plateau
$\chid(t,\tw) \approx 2/\sqrt{\pi \tw}$ for $\dt \ll \tw$ according to
$\chid(t,\tw) \approx 4c(\tw) (1 - p_\mathrm{r}(t-\tw))$. This
crossover is clear from the spin-chain dynamics: the perturbation
associated with $\chid(t,\tw)$ is $\delta H = - h s_i s_{i+1}$ which
simply increases the coupling between sites $i$, $i+1$. This enforces
alignment of the spins $s_i$ and $s_{i+1}$ and hence increases
$\langle s_i(t) s_{i+1}(t) \rangle$ on a microscopic time scale.  In
the aging limit the leading term in the integrated response is just
$\chid(t,\tw) \sim 2/\sqrt{\pi t}$ which connects to the plateau
$\chid(t,\tw) \approx 2/\sqrt{\pi \tw}$ for $\dt \ll \tw$ but
eventually decreases as $\chid(t,\tw) \approx 2/\sqrt{\pi \dt}$ for
$\dt \gg \tw$.
 
\begin{figure}[htb]
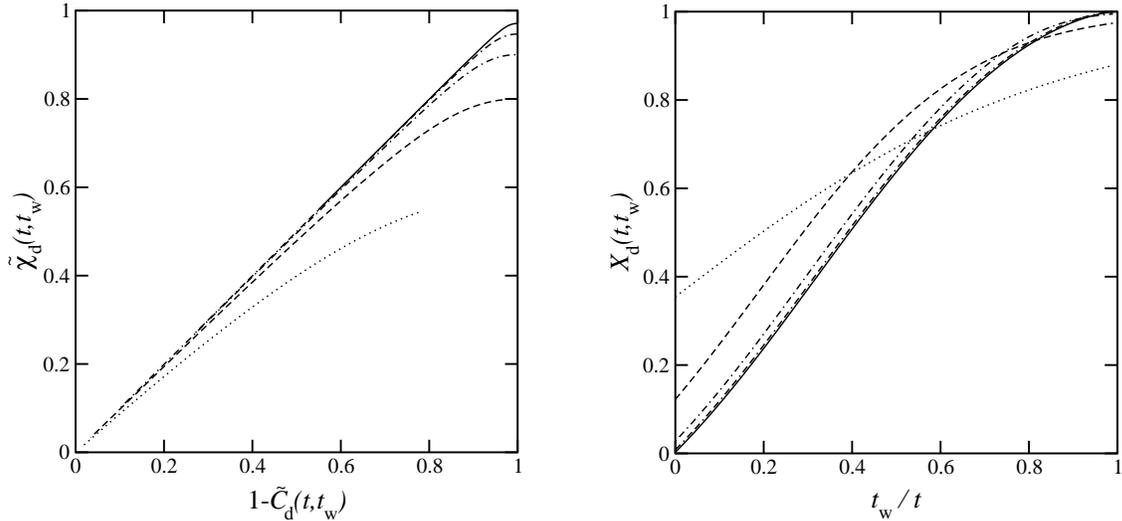
 
  \begin{picture}(15,7) \put(0,-0.2){\epsfig{file=DefectFDT.eps
    ,scale=0.4,clip}} \put(8,-0.2){\epsfig{file=DefectX.eps
    ,scale=0.4,clip}} \end{picture} \caption{\label{fig:defectFDplot}
    Normalized FD-plot (left) and the corresponding FDR versus $\tw/t$
    (right) for the defect autocorrelation and susceptibility. In both
    plots $t$ is kept fixed, giving a one-to-one correspondence of
    the curves, and varies over the range $t=10^0$ (dotted), $10^{1/2},
    10^1,10^{3/2},10^2$ (solid). The curves of
    $\Xd(t,\tw)$ for $t=10^{3/2},10^2$ are almost indistinguishable
    and very close to the limit curve (\ref{equ:Xd}).}
\end{figure} 
For constructing an FD plot (Fig.~\ref{fig:defectFDplot}) we are
interested in keeping $t$ fixed and varying $\tw$ between $0$ and $t$;
the functions $\cord(t,\tw)$ and $\chid(t,\tw)$ are then monotonic in
$\tw$. In fact the exact expressions (\ref{equ:cordrndw}),
(\ref{equ:chidrndw}) satisfy $\cord(t,t) - \cord(t,\tw) = \chid(t,\tw)
+ 4[c^2((t+\tw)/2)-c^2(t)]$. Dividing this relation by the equal time
value $\cord(t,t)$ yields the relevant normalized quantities
\begin{equation} 
  1-\tilde{C}_\mathrm{d}(t,\tw) = \tilde{\chi}_\mathrm{d}(t,\tw) +  
  \frac{c^2\left(\frac{t+\tw}{2}\right)-c^2(t)}{c(t)(1-c(t))} =  
  \tilde{\chi}_\mathrm{d}(t,\tw) + \mathcal{O} \left( \frac{1}{\sqrt{t}}  
  \, \frac{t-\tw}{t+\tw} \right) .
  \label{equ:dFDT} 
\end{equation} 
In the limit $t \to \infty$ the extra term in (\ref{equ:dFDT}) vanishes  
and we get $1-\tilde{C}_\mathrm{d}(t,\tw) = \tilde{\chi}_\mathrm{d}(t,\tw)$  
for all $0 \leq \tw/t \leq 1$. This, however, does not imply that  
equilibrium FDT holds, i.e. $\Xd(t,\tw) = 1$. In fact, working out  
$(\partial/\partial \tw)\cord(t,\tw)$ and $\resd(t,\tw)$ from  
(\ref{equ:cordrndw}), (\ref{equ:chidrndw}) and expanding their ratio  
in the aging limit gives 
\begin{equation} 
  \Xd(t,\tw) \sim \frac{\tw (t+\tw)}{\tw (t+\tw)+(t-\tw)\sqrt{t^2-\tw^2}}. 
  \label{equ:Xd} 
\end{equation} 
The FDR (\ref{equ:Xd}) is a function of the time ratio $\tw/t$ and
crosses over from $\Xd(t,\tw) = 1$ for $\tw/t \to 1$ to $\Xd(t,\tw)
= \Xd^\infty = 0$ for $\tw/t \to 0$ (Fig.~\ref{fig:defectFDplot}).
This seemingly paradoxical result can easily be explained in terms of
the expansions given above. In the regime $\dt \ll t$ (which is
equivalent to $\dt \ll \tw$, as considered before) we have, up to
subleading corrections for $t \to \infty$, $\cord(t,\tw) \approx 4c(t)
p_\mathrm{r}(t-\tw)$ and $\chid(t,\tw) \approx 4c(t)
(1-p_\mathrm{r}(t-\tw))$. So equilibrium FDT indeed holds in this
regime and the DLPA process is, to leading order, just an ensemble of
independent random walks. Now recall that $p_\mathrm{r}(\dt) \approx
1/\sqrt{2 \pi \dt}$ for $\dt \gg 1$. So at the point where this
approximation breaks down, $\dt \approx t$, the value of, e.g.,
$\cord(t,\tw)$ decreases to an arbitrary small fraction of
$\cord(t,t)$ as $t$ increases. This leads to a straight line segment
which eventually covers the whole of the (normalized) FD plot while
the size of the non-trivial region shrinks as $1/\sqrt{t}$. In the
latter part, for $\tilde{C}_\mathrm{d}(t,\tw) \ll 1/\sqrt{\pi t}$, one
has from the aging expansions of $\cord(t,\tw)$ and $\chid(t,\tw)$
\begin{equation} 
  1-\tilde{\chi}_\mathrm{d}(t,\tw) \approx \frac{1}{\sqrt{\pi t}} +  
  \frac{\sqrt{\pi t}}{2} \tilde{C}^2_\mathrm{d}(t,\tw) .
  \label{equ:dFDTapx} 
\end{equation} 
Hence the FD plot indeed turns horizontal as $\tilde{C}_\mathrm{d}(t,\tw)$ 
approaches zero, consistent with (\ref{equ:Xd}). In summary, an  
FD plot is not the appropriate representation for the FDT violation  
measured by the defect autocorrelation and response. A plot of the FDR  
as a function of  
$\tw/t$, however, converges to the non-trivial limit curve 
 given by (\ref{equ:Xd}) as times 
diverge, see  
Fig.~\ref{fig:defectFDplot}.

\subsubsection{Uniform field: Coherent functions} 
\label{4spin1de} 

For uniform covariances $q_n=1$ the defect observable $O_\mathrm{d}$ 
is equivalent to the total energy of the system. According to 
(\ref{equ:corchiOfourier}) we have $\core(t,\tw)=C(0;t,\tw)$
which may be simplified to give
\begin{equation} 
  \core (t,\tw) = 4 \me^{-2 t} [\In_0+\In_1](2 t) -  
    \me^{-2 (t+\tw)} [3 \In_0 + 4 \In_1 + \In_2](2(t+\tw))  .
  \label{equ:core} 
\end{equation} 
This result again has an analog in the associated DLPA process, where it
describes the normalized two-time correlation of the total number of
particles $\mathcal{N}$, $\core(t,\tw) = (4/N)
\left[ \langle \mathcal{N}(t) \mathcal{N}(\tw) \rangle -\langle
\mathcal{N}(t) \rangle \langle \mathcal{N}(\tw)\rangle \right]$. 
A result similar to (\ref{equ:core}) was given in~\cite{schuetz}, for
initial conditions corresponding formally to equilibrium at inverse
temperature $1/T=-\infty$. Up to a factor 
of 4 which appears to be missing in~\cite{schuetz}, it coincides  
with~(\ref{equ:core}) for large $\tw$, where one finds the simple scaling
form $\core(t,\tw) \sim 4/\sqrt{\pi} ( 1/\sqrt{t}-1/\sqrt{t+\tw}
)$. At {\em equal} times, fluctuations in the energy follow as
$\core(\tw,\tw) \sim (2-\sqrt{2}) \cord(\tw,\tw)$. This shows that in
$\core(\tw,\tw)=\sum_n \cord[n](\tw,\tw)$ the non-local ($n\neq0$)
terms make a contribution $-(\sqrt{2}-1)\cord(\tw,\tw)$, of the same
order as the local term but with opposite sign.  For $\dt \ll \tw$ the
two-time correlation $\core(t,\tw) \approx \core(\tw,\tw)$ has a
plateau but it decreases as $\core(t,\tw) \approx 2 \tw/(\dt
\sqrt{\pi \dt})$ when $\dt \gg \tw$.

By setting $k=0$ in the Fourier transform (\ref{equ:chidfourier}) we
find that $\chie(t,\tw) = \chi(0;t,\tw) \equiv 0$ at all times.  This
is for the simple reason that the perturbation is proportional to the
Hamiltonian and therefore just rescales the temperature, which
obviously has no effect in the $T \to 0$ limit considered here.  We
note that $\chie(t,\tw) = \sum_n \chid[n](t,\tw) = 0$ implies that the
sum over all cross-susceptibilities ($n\neq0$) exactly balances the
local susceptibility $\chid(t,\tw)\equiv\chid[0](t,\tw)$.

An FD-plot for the energy is obviously just a horizontal line and the
corresponding FDR is $\Xe(t,\tw)=\Xe^\infty=\Xd^\infty=0$.  This
matches our findings in Sec.~\ref{2spin1d} in the sense that the
FD-plot for the coherent observable is a straight line whose
slope is the $X^\infty$ of the incoherent observable.

\subsubsection{Short-range correlated field} 
\label{4spin1dL} 

We have seen above that for local defect observables the FD plot is
not appropriate for determining FDT violation effects, since it
converges to a straight line in the aging limit. It turns out that the
same holds for defect observables defined by short-range correlated
fields. To see this, recall from~(\ref{equ:corchiO}) that e.g.\ the
correlation function $C(t,\tw)$ of the observable is a weighted sum of
the non-local defect correlations, and focus on the regime
$\dt=\mathcal{O}(1)$ that dominates the FD plot for large $\tw$ or $t$.
From~(\ref{equ:cordn}), one then easily shows that whenever a
non-local term with given $n\neq 0$ is of the same order as the local
contribution $\cord(t,\tw)\equiv \cord[0](t,\tw)\sim 2/\sqrt{\pi \tw}
\me^{-\dt} \In_0(\dt)$, it can be written as
\begin{equation}
\cord[n](t,\tw) \sim \frac{2}{\sqrt{\pi\tw}}{\rm e}^{-\dt}\In_n(\dt).
\label{corr:fixed_n}
\end{equation}
In the same regime the expression for the non-local susceptibility
$\chid[n](t,\tw)$ is identical apart from a minus sign. This
shows that, whatever the short-ranged field correlations $q_n$, the
FD plot of $\chi(t,\tw)$ vs $C(t,\tw)$ for the observable considered
becomes trivial for long times, just as in the case
$q_n=\delta_{n,0}$. We therefore focus on the FDR in the following,
which requires analysis of $(\partial/\partial \tw)C(t,\tw)$ and the
response $R(t,\tw)=-(\partial/\partial\tw)\chi(t,\tw)$ and should
become non-trival in the aging limit.

By analogy with the results presented in Sec.~\ref{2spin1dL} for spin
observables, we will show that the FDR becomes identical to that for
the incoherent functions in the aging limit. The procedure is again to
prove that the Fourier transforms of the defect functions
$(\partial/\partial \tw)C(k;t,\tw)$ and $R(k;t,\tw)$ are 
representations of $\tilde{\delta}$ in the
aging-limit and with appropriate normalization.

The expressions (\ref{equ:cordfourier}) for $C(k;t,\tw)$ and the one
that follows from (\ref{equ:chidfourier}) for $R(k;t,\tw)$ are rather
complicated and it is a priori not clear how they behave as times
diverge. Asymptotic expansions in the aging limit $t,\tw \to \infty$
with $\epsilon \leq \tw/t \leq 1-\delta$ fixed ($\epsilon, \delta >
0$) and $|k| \leq K$ where $K=c/\sqrt{\tw}$ ($c > 0$ arbitrarily large 
but finite),
however, capture the relevant features of $C(k;t,\tw)$, $R(k;t,\tw)$
and have a considerably simpler form:
\begin{eqnarray} 
  C(k;t,\tw)  &\sim&  \frac{4}{\sqrt{\pi}} \Big\{  
    \frac{1}{\sqrt{t}} \, \me^{-k^2 (t^2-\tw^2)/(4t)} - 
    \frac{1}{\sqrt{t+\tw}} \, \me^{-k^2 (t+\tw)/4}  
    \Big\} \label{equ:cordfourierapx} \\[1ex] 
  &+&  2 k \me^{-k^2 (t+\tw)/2} \left\{  
    \erfi\left(k {\textstyle \frac{t+\tw}{2 \sqrt{t}} } \right) -  
    \erfi\left(k {\textstyle \frac{\sqrt{t+\tw}}{2} } \right) \right\} , 
    \nonumber \\[1ex] 
  R(k;t,\tw) &\sim& \frac{1}{\sqrt{\pi t}} 
    \left(\frac{\tw}{t}\right) k^2 \, \me^{-k^2 (t^2-\tw^2)/(4t)} .
  \label{equ:resdfourierapx} 
\end{eqnarray} 
$\erfi(x)$ is the error function with imaginary argument:
$\erfi(x)=(1/i)\mathrm{Erf}(ix)$~\cite{mathbook}.  Note that the
arguments of all exponentials and the $\erfi$'s are of
$\mathcal{O}(1)$ if $\tw/t$ and $k$ are in the specified range. For
$|k|$ larger than $\mathcal{O}(1/\sqrt{\tw})$ the 
results~(\ref{equ:cordfourierapx}), (\ref{equ:resdfourierapx})
do not apply.

In (\ref{equ:cordfourierapx}) the growth of $\erfi(x) \sim
\me^{x^2}/(\sqrt{\pi} x)$ is over-compensated by the exponential
prefactor and so we can make $C(K;t,\tw)$ arbitrarily small by 
increasing $c$.
For larger $k$, $|k|>K$, 
the values of $C(k;t,\tw)$ as given by (\ref{equ:cordfourier})
also turn out to be insignificant.
Therefore $C(k;t,\tw)$ develops an infinitely sharp
peak of width $\mathcal{O}(1/\sqrt{\tw})$ 
at $k=0$ in the aging limit and becomes a realization of
$\tilde{\delta}(k)$ when normalized by $\cord(t,\tw)$, in analogy with
(\ref{equ:spincorchidelta}). Differentiating
(\ref{equ:cordfourierapx}) w.r.t. $\tw$ turns out to reproduce the
rigorous expansion for $(\partial / \partial \tw) C(k;t,\tw)$ and
similar arguments apply. Hence $C(t,\tw) \sim N_\mathrm{c}
\cord(t,\tw)$ and $(\partial / \partial \tw) C(t,\tw) \sim
N_\mathrm{c} (\partial / \partial \tw) \cord(t,\tw)$ for any
short-range correlated field with $N_\mathrm{c} \neq 0$.

The expansion (\ref{equ:resdfourierapx}) for the response function
$R(k;t,\tw)$ also peaks sharply in the region $|k| \leq K$ near $k=0$;
it follows from (\ref{equ:chidfourier}) that outside this $k$-range
$R(k;t,\tw)$ is insignificantly small again. We have,
however, $R(0;t,\tw)=0$ at all times. Nevertheless, $\resd(t,\tw)$
yields normalization and the ratio of both vanishes in the aging-limit
for any $k \neq 0$ (modulo $2\pi$). So $R(k;t,\tw)$ becomes a realization
of $\tilde{\delta}(k)$ when normalized by $\resd(t,\tw)$, i.e.~two
infinitely sharp peaks at $0^+$ and $0^-$. Therefore $R(t,\tw) \sim
N_\mathrm{c} \resd(t,\tw)$ for any short-range correlated field with
$N_\mathrm{c} \neq 0$.

Since $R(t,\tw) \sim N_\mathrm{c} \resd(t,\tw)$ and $(\partial /
\partial \tw) C(t,\tw) \sim N_\mathrm{c} (\partial / \partial \tw)
\cord(t,\tw)$, any defect observable $O_\mathrm{d}$ with short-range
correlated fields $\epsilon_i$ and $N_\mathrm{c} \neq 0$ ultimately
gives the same FDR as the incoherent functions. The scaling of the
peaks in (\ref{equ:cordfourierapx}), (\ref{equ:resdfourierapx})
implies associated time-dependent length scales in real space. As in
the spin observable case the FDR will thus display a crossover (see
Fig.~\ref{fig:defectpowerlimit}) when these length scales become
comparable with the length over which the fields $\epsilon_i$ are
correlated. We note finally that, in contrast to the response
$\resd[n](t,\tw)$ discussed above, the integrated response or
susceptibility $\chid[n](t,\tw)$ displays somewhat unusual behaviour;
e.g.\ the local value $\chid[0](t,\tw)$ dominates the non-local terms
for all times, so that to leading order there is no real-space length
scale associated with the defect susceptibility.  One also finds
non-trivial features in the FD plots and FDRs for the
cross-correlations $\cord[n](t,\tw)$ and susceptibilities
$\chid[n](t,\tw)$~\cite{Peter_thesis}.  However, in the aging-limit
and for any fixed $n\neq 0$ the FDR for the local observable ($n=0$)
is recovered as in the spin observable case.

\subsubsection{Infinite-range correlated field} 
\label{4spin1dP} 
 
We next consider the FDR for observables defined by infinite-range
correlated fields. As for short-range correlated fields, we will not
discuss the integrated quantities $C(t,\tw)$ and $\chi(t,\tw)$ in
detail. One finds again that these give a trivial FD plot for long
times, although the argument for this is somewhat more subtle than
in~(\ref{corr:fixed_n}) because one needs to consider an infinite
range of distances $n$.

\begin{figure}[htb]
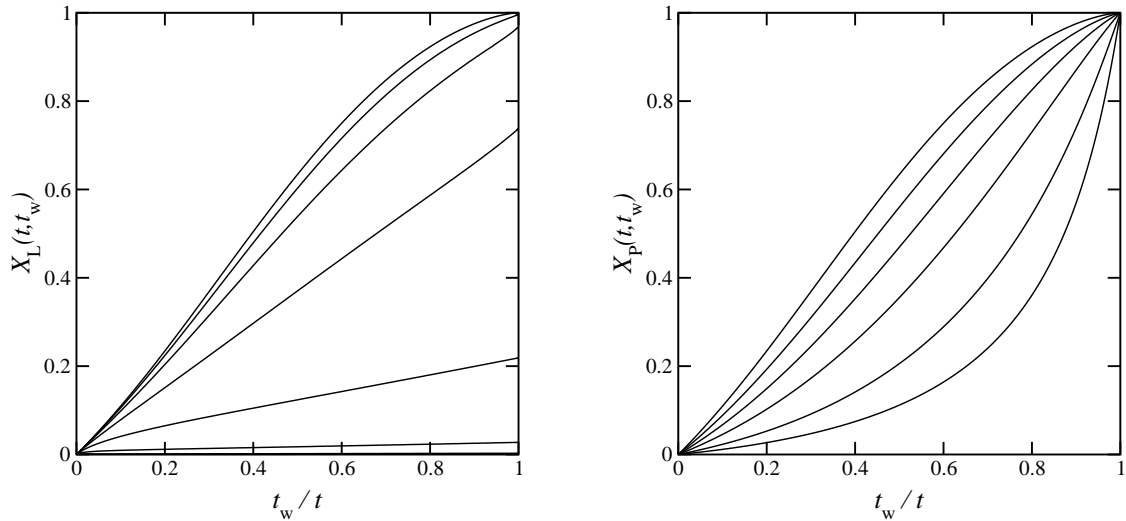
 
  \begin{picture}(15,7) 
    \put(0,0){\epsfig{file=DefectLorenz100X.eps    ,scale=0.4,clip}} 
    \put(8,0){\epsfig{file=DefectPowerLimitX.eps,scale=0.4,clip}} 
  \end{picture} 
  \caption{\label{fig:defectpowerlimit} Left: Time evolution of the
defect-observable FDR for 
    a random field with Lorentzian covariances (\ref{equ:lorenzQ}) and 
    $a=10^2$, correlated over $N_c=a \pi \coth a \pi\approx a\pi$ sites.
    For each curve $\tw$ is kept fixed, varying over the range
$\tw=10^0,\ldots 
    10^6,\infty$ from bottom to top. The curves for 
    $\tw=10^0, 10^1, 10^2$ are flat; in this regime the observable is
effectively identical to the energy. For $\tw=10^3, 10^4, 
    10^5,10^6$ we see the crossover to the limit curve for $\tw \to \infty$ 
    given by (\ref{equ:Xd}) and corresponding to the incoherent observable.
    Right: Limit 
    curves of the FDR versus $\tw/t$ for $t,\tw \to \infty$. 
    From bottom to top these correspond 
    to power-law covariances with exponent $\alpha=0.1, 0.2, 0.4, 0.6, 0.8$. 
    The top curve shows the random-field limit $\alpha \to 1$ for
    power-laws, and also applies to any short-range correlated field 
    Sec.~\ref{4spin1dL} with $N_\mathrm{c} \neq 0$ or, in the extreme 
    case, the incoherent functions of Sec.~\ref{4spin1dd}.}
\end{figure}

The exact expressions for the two-time correlation functions and
susceptibilities for defect observables with power-law covariances
follow from~(\ref{equ:corchiOfourier}) by substitution of
$C(k;t,\tw)$, Eq.~(\ref{equ:cordfourier}), $\chi(k;t,\tw)$,
Eq.~(\ref{equ:chidfourier}), and $q_{\power}(k)$,
Eq.~(\ref{equ:powerQ}). The response $R(t,\tw)$ is then obtained from
$\chi(t,\tw)$ by $R(t,\tw)=-(\partial/\partial\tw)\chi(t,\tw)$ as
usual.  The resulting equations are rather bulky and too complex for a
meaningful discussion. So we immediately turn to the aging limit,
where we can use the following, asymptotically exact, approximations.
Firstly we replace the exact expressions in (\ref{equ:corchiOfourier})
for $C(k;t,\tw)$, $R(k;t,\tw)$ by (\ref{equ:cordfourierapx}),
(\ref{equ:resdfourierapx}). Although (\ref{equ:cordfourierapx}),
(\ref{equ:resdfourierapx}) do not hold outside the range $k\in
[-K,+K]$, the contributions to the $k$-integrals are subleading.
Secondly, as the integrands have infinitely sharp peaks at $k=0$ in the
aging limit, we may replace $q_{\power}(k)$ by the leading term of its
expansion at $k=0$, i.e. replace $\sin(k/2)$ by $k/2$ in
(\ref{equ:powerQ}).  This in turn allows us to extend the limits of
integration in (\ref{equ:corchiOfourier}) from $-\pi,+\pi$ to
$-\infty,+\infty$, whereby we again just accumulate subleading
errors. Having made these approximations, which still yield
asymptotically exact results, the $k$ integrations can be evaluated
and we get
\begin{eqnarray} 
  \cord[\power](t,\tw) &\sim& \frac{2}{\pi}  
    \Gamma\left({\textstyle \frac{1+\alpha}{2}}\right) \left\{  
    t^{\frac{\alpha-1}{2}} \left( t^2-\tw^2 \right)^{-\frac{\alpha}{2}} 
    - (t+\tw)^{-\frac{1+\alpha}{2}} \mathrm{F}\left({\textstyle  
    \alpha; \frac{t+\tw}{2t}} \right) 
    \right\}, \label{equ:cordpowerapx} \\ 
  \resd[\power](t,\tw) &\sim& \frac{2}{\pi} \alpha \,
    \Gamma\left({\textstyle \frac{1+\alpha}{2} } \right)  
    \tw t^{\frac{\alpha-1}{2}}  
\left( t^2-\tw^2 \right)^{-(1+\frac{\alpha}{2})} ,
    \label{equ:resdpowerapx} 
\end{eqnarray} 
where we have introduced the short-hand 
\begin{equation} 
  \mathrm{F}\left({\alpha; \textstyle\frac{t+\tw}{2t}}\right) =  1 -  
    (1-\alpha) \, 2^{-\frac{1+\alpha}{2}} \left(  
    \mathrm{B}\left({\textstyle 
    \frac{1}{2},1-\frac{\alpha}{2};\frac{t+\tw}{2t} }\right) -  
    \mathrm{B} 
    \left({\textstyle \frac{1}{2},1-\frac{\alpha}{2}; 
\frac{1}{2} }\right)\right) .
\end{equation} 
In the limit $\alpha \to 1$, Eqs.~(\ref{equ:cordpowerapx}),
(\ref{equ:resdpowerapx}) reduce to the asymptotic expansions of the
incoherent functions $\cord(t,\tw)$, $\resd(t,\tw)$ while $\alpha \to
0$ gives the asymptotic expansions of the coherent ones,
i.e. $\core(t,\tw)$ from (\ref{equ:cordpowerapx}) and $\rese(t,\tw)=0$
from (\ref{equ:resdpowerapx}).  So the power-law covariances
(\ref{equ:powerQ}) again allow us to interpolate between local and
global observables. For intermediate exponents $0<\alpha<1$ the two-time
correlations in $O_\mathrm{d}$ decrease as $\tw^{-1/2}
\dt^{-\alpha/2}$ in the regime $1 \ll \dt \ll \tw$ and cross over to
$\tw \dt^{-(3+\alpha)/2}$ for $1 \ll \tw \ll \dt$.
The response
$\resd[\power](t,\tw)$ behaves as $\tw^{-1/2} \dt^{-(2+\alpha)/2}$ for
$1 \ll \dt \ll \tw$ and $\tw \dt^{-(5+\alpha)/2}$ for $1 \ll \tw \ll
\dt$.  An aging expansion for the FDR again gives non-trivial curves.
The derivative
$(\partial/\partial \tw)\cord[\power](t,\tw)$ follows correctly by
differentiating the expansion (\ref{equ:cordpowerapx}) which, together
with (\ref{equ:resdpowerapx}) yields
\begin{equation} 
  \Xd[\power](t,\tw) \sim \left\{ 1 + \frac{t-\tw}{\tw} \left[  
    \frac{1-\alpha}{2 \alpha} + 
    \frac{1+\alpha}{2 \alpha} \sqrt{ \frac{t}{t+\tw} } 
    \left(\frac{t-\tw}{t}\right)^{\frac{\alpha}{2}} 
    \mathrm{F}\left({\textstyle \alpha; \frac{t+\tw}{2t} }\right)   
    \right] \right\}^{-1} .
  \label{equ:Xdpowerapx} 
\end{equation} 
The FDR 
$\Xd[\power](t,\tw)$ is a function of $\tw/t$ only and interpolates
between the FDR (\ref{equ:Xd}) for the local defect observables
($\alpha \to 1$) and $\Xe(t,\tw)=0$ for the energy ($\alpha \to
0$). For any power $0<\alpha<1$ (\ref{equ:Xdpowerapx}) crosses over
from $\Xd[\power](t,\tw) = 1$ for $\dt \ll \tw$ to $\Xd[\power](t,\tw)
= \Xd[\power]^\infty=0$ for $\dt \gg \tw$ (see
Fig.~\ref{fig:defectpowerlimit}).

\subsubsection{Harmonically correlated fields and $X^\infty$} 
\label{4spin1dX}
 
In contrast to spin observables it appears that for defect observables
$O_\mathrm{d}$ we generically find $X^\infty=0$.  To prove this claim
we may again follow the approach presented in
Sec.~\ref{2spin1dX}. Introducing an FDR for defect Fourier modes
$X(k;t,\tw)$ according to (\ref{equ:FDTsfourier}) based on the
two-time defect correlation function (\ref{equ:cordfourier}) and
susceptibility (\ref{equ:chidfourier}) allows us to write the FDR for
any defect observable in the form (\ref{equ:XsQ}). The full expression
for $X(k;t,\tw)$ is rather complicated and, in contrast to (\ref{equ:Xsk}),
retains a non-trivial dependence on $k$, $t$ and $\tw$. The only
general features are $X(0;t,\tw)=0$, since $X(0;t,\tw)=\Xe(t,\tw)=0$,
and $X(\pm \pi;t,\tw)=1+\mathcal{O}(\sqrt{\tw} \me^{-4\tw})$ being
independent of $t$ and close to one for $\tw \gg 1$.  For intermediate
values $0<|k|<\pi$ the FDR $X(k;t,\tw)$ can in fact take arbitrarily
large values for appropriate $\tw$ and $t$.  To repeat the argument of
Sec.~\ref{2spin1dX}, however, we just have to be able to take the
limit $t \to \infty$ for fixed and finite $\tw$. A scaling analysis of
$(\partial/\partial \tw)C(k;t,\tw)$ as obtained from
(\ref{equ:cordfourier}) shows that this quantity develops an
infinitely sharp peak at $k=0$. Hence the normalized distribution of
$q(k) (\partial/\partial \tw) C(k;t,\tw)$ over $-\pi \leq k \leq \pi$
becomes, for sufficiently well behaved functions $q(k)$, a realization
of $\tilde{\delta}(k)$ as $t \to \infty$. Remarkably, the FDR for
defects $X(k;t,\tw)$ approaches a simple, smooth function on $\pi < k
< \pi$ in the same limit
\begin{equation} 
  \lim_{t \rightarrow \infty} X(k;t,\tw) =  
    2 \sin^2 \frac{k}{4}  
    \left(1+4\tw \cos^2 \frac{k}{4} \right) 
    \left(2-\me^{-4\tw \sin^2 \frac{k}{4} }\right)^{-1}.
  \label{equ:Xdtinfty} 
\end{equation} 
Together, these two facts imply that out of the spectrum of FDRs
$X(k;t,\tw)$ for Fourier modes $k$, the long time limit $t \to \infty$
again selects the contributions associated with infinite length scales
($k=0$). These are, in the limit, given by (\ref{equ:Xdtinfty}) and
equal to zero. The FDR (\ref{equ:XsQ}) for defect observables
$X(t,\tw) \to 0$ thus vanishes as $t \to \infty$ regardless of the
choice of $q(k)$, except in pathological cases as discussed in
Sec.~\ref{2spin1dX}. Note that because (\ref{equ:Xdtinfty}) for $k=0$
gives a vanishing result for any $\tw$, one in fact has
$\lim_{t\to\infty} X(t,\tw)=X^\infty=0$, without needing to take
$\tw\to\infty$.

We note finally that the behaviour at short wavelengths is rather more
complex for defect observables than for spins. In particular, even for
what one might expect to be `equilibrated' wavelengths, $k^2\tw\gg 1$,
it is not true that $X(k;t,\tw)\approx 1$ for all times $t$, and
$X$ deviates significantly from this simple value for large time
differences $\dt\gg\tw$ as can be seen from (\ref{equ:Xdtinfty}).

\subsection{Physical discussion} 
\label{physdis}
 
We saw above that apart from pathological exceptions all spin and
defect observables give {\it identical} values for the asymptotic FDR
$X^\infty$, with $X^\infty=1/2$ for spin observables and $X^\infty=0$
for defect observables. These slopes are most easily read off from the
FDT plots for the coherent observables (magnetization and energy,
respectively) which become straight lines in the long-time limit.

It is natural to ask how these results would extend to observables
other than those we have considered, such as $O=\sum_i
\epsilon_i s_i s_{i+2}$ which involves spin pairs at distance two.  We
have worked out explicitly the FD properties based on the general
solutions given in \cite{peterpeter2} for the coherent and incoherent
versions for this observable \cite{Peter_thesis}; one finds that they are, 
up to
subdominant corrections, identical to those for $O=2\sum_i \epsilon_i
s_i s_{i+1}$. The physical interpretation is simple: $s_i s_{i+2}=-1$
if there is exactly one domain wall between spins $i$ and $i+2$, while
$s_i s_{i+2}=1$ if there is no domain wall or if there are two. The
last alternative, however, is suppressed in the aging limit where
typical distances between domain walls scale as $\sqrt{\tw}$, and so
$s_i s_{i+2}\approx s_{i}s_{i+1} + s_{i+1}s_{i+2} - 1$. For the
coherent observable ($q_n=[\epsilon_i\epsilon_{i+n}]=1$), this
directly explains our observation; for the incoherent version
($q_n=\delta_{n,0}$) it follows from the fact that the correlations of
$s_{i}s_{i+1}$ and $s_{i+1}s_{i+2}$ are identical to the
autocorrelations of $s_{i}s_{i+1}$ in the aging limit.

By a similar reasoning we can now predict the FD behaviour of
higher-order observables of the form
\begin{equation} 
  O^{(k)}_{\boldsymbol{j}}=\sum\limits_i \epsilon_i \prod\limits_{\eta=1}^k  
  s_{i+j_\eta} ,
  \label{equ:observable}
\end{equation} 
with $k\geq 3$ and $j_1=0$; $j_2,\ldots, j_k$ specify the relative
displacements of the spins in the $k$-th order products. For
even $k$, each term $\prod_{\eta} s_{i+j_\eta}$ again has a sign
depending on the number of domain walls between spins $i$ and
$i+j_k$. In the aging limit, configurations where more than one domain
wall occurs can be neglected, so that we can replace the product by
$s_{i} s_{i+j_k}$. By the same argument as above, this is essentially
equivalent to $j_k\times s_i s_{i+1}$ and so should again give
$X^\infty=0$.

For odd $k$, on the other hand, the sign of $\prod_{\eta}
s_{i+j_\eta}$ is essentially determined by the sign of the domain
which the spin $s_i$ finds itself in. The leading contribution is now
given by configurations with no domain walls between $s_i$ and
$s_{i+j_k}$. Configurations with at least one domain wall are again
suppressed in the aging limit. We can therefore replace the product
simply by $s_i$ to leading order, giving an asymptotic FDR of
$X^\infty=1/2$ as for genuine first order spin observables. 

The fact that observables of even and odd order behave in different
ways can also be motivated mathematically from the hierarchy of the
equations obeyed by the multi-spin correlation
functions where the even and odd orders $k$
turn out to decouple completely~\cite{peterpeter2}. 
This is a peculiarity of the
one-dimensional Ising model, whereas in the generic case one would
expect all levels of the hierarchy to couple to each other, resulting
in a unique value of $X^\infty$. We indeed find strong evidence for
this in the two-dimensional case below.

From a more physical point of view, the existence of two different
values of $X^\infty$ could be related to the fact that in the
one-dimensional chain at $T=0$ one has both a critical point and
an ordered phase. The result $X^\infty=0$ for defect observables could
thus be related to the ordinary results for coarsening in $d\geq 2$
after a quench to an ordered phase, while $X^\infty=1/2$ for the
spins would reflect the critical aspects of coarsening at $T=T_c=0$.

It might 
be interesting -- though rather complicated -- to use the methods 
described above and in
Ref.~\cite{peterpeter2} to study higher order observables 
different from (\ref{equ:observable}) for which the above leading 
order approximations do not apply, 
for example, $1 - s_{i}s_{i+1} - s_{i+1}s_{i+2} + 
s_i s_{i+2}$, which corresponds to a quadratic operator in
bond variables, $4 \, b_i b_{i+1}$. We are currently exploring this issue.

To recap, the central result of this section is that (almost) all
observables of the  
form (\ref{equ:observable}) 
interpolate between an equilibrium like behaviour with $X=1$ and an
asymptotic FDR $X^{\infty}$. The latter are given by the values
of $X(k \to 0)$, as was argued in
\cite{gambas1,gambas2,gambas3}. We have shown that the most efficient
way of extracting $X^{\infty}$ is by studying coherent
functions.
These results motivate the following section where the 2$d$
Ising model is studied at criticality. 

\section{The 2$\boldsymbol{d}$ Ising model} 
\label{section2d} 
 
In this section, we report on numerical simulations of the $2d$ Ising
model.  It is defined by the Hamiltonian
\begin{equation} 
\mathcal{H} = - \sum_{\langle i,j \rangle} s_i s_j, 
\end{equation} 
where the $s_i$
$(i=1,\cdots,N )$ are $N$ Ising spins located on
the sites of a square lattice with periodic boundary conditions and
linear size $L$; the sum is over nearest neighbor pairs.  We perform
Monte-Carlo simulations using a standard Metropolis algorithm where
the spins are randomly updated. One Monte Carlo step represents $N$
attempts to flip a spin.
 
The system is prepared in a random state, corresponding to an infinite
initial temperature.  It is then quenched at $t=0$ to the critical
temperature $T_c = \frac{1}{2} \ln(1+\sqrt{2})$.  As stated in
Sec.~\ref{ferromotiv}, we focus on the four natural FD relations for
the Ising model, constructed from the coherent and incoherent
dynamical functions of spin and defect observables.  The system size
we use is different for coherent and incoherent objects.  Incoherent
objects reflect the behaviour of individual spins or defects, and
simulating a very large system is advantageous in that it makes an
average over many initial conditions unnecessary.  Coherent objects
like $\core(t,\tw)$ or $\corm(t,\tw)$, on the other hand, have an
amplitude of order $1/N$.  One should thus simulate many initial
conditions of the smallest possible system, with the opposite
constraint that the system has to be out of equilibrium even for the
largest simulated time scale, $t_{\rm simu}$, giving the condition
$\xi(t_{\rm simu}) \ll L$.  Our results are obtained with $t_{\rm
simu} = 10^5$, $L=300$ for coherent functions, and $L=500$ for
incoherent ones.  Only a few samples over initial conditions are
necessary for incoherent correlation functions, while 1000 initial
conditions were sampled for coherent ones.  This is also the number of
realizations necessary to get the four susceptibilities we have
computed.
 
We now describe our results, starting with spin observables and then
turn to defect observables.
 
\subsection{Spin observables} 
 
\begin{figure} 
  \psfig{file=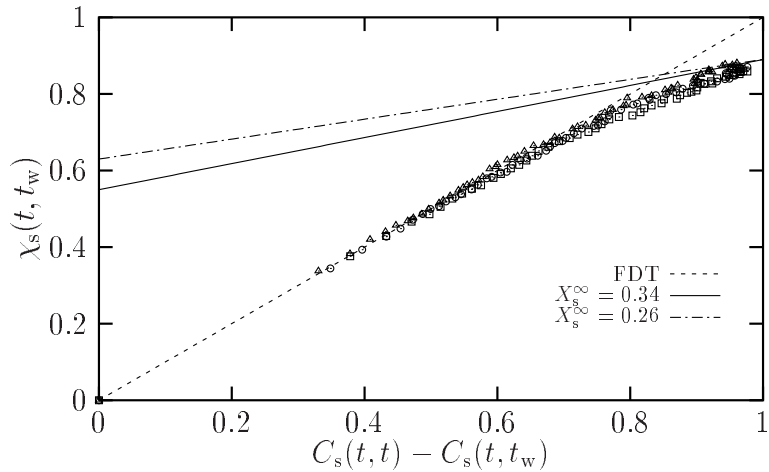,width=11cm} 
  \caption{FD plots for spin autocorrelation and response. 
  Three waiting times, $\tw=43$, 179 and 460 are represented  
  by squares, circles and triangles, respectively.  
  The dashed line with slope one shows the equilibrium FDT. 
  The full and dash-dotted lines have slopes 
  $\Xs^\infty=0.34$ and $\Xs^\infty=0.26$, respectively; these are 
  discussed in the text.} 
  \label{2dspin} 
\end{figure} 
 
The two-time scaling behavior of the incoherent spin functions
$\cors(t,\tw)$ and $\chis(t,\tw)$ has been the subject of a number of
publications, as described in Sec.~\ref{ferrocrit}.  We refer to the
references cited there and directly present in Fig.~\ref{2dspin} the
FD plot for the spin autocorrelation and susceptibility.  A very
similar FD plot has been reported in Ref.~\cite{golu2}, although a
somewhat different susceptibility $\int_0^{\tw} \upd \tau \,
\ress(t,\tau)$ was plotted there, so that the FD plot looks reversed
compared to Fig.~\ref{2dspin}.  Otherwise, we find the features
anticipated in Sec.~\ref{ferrocrit}.  The FD plot is characterized by
an initial part which follows the equilibrium FDT, corresponding to
short, equilibrated length scales.  For larger time differences, the
FD plot deviates from the FDT in a non-trivial manner due to the
non-equilibrated fluctuations at small wavevectors.  In the limit of
large time differences the FD plot has a nonzero slope $\Xs^\infty$,
in contrast to the zero slope obtained below the critical 
point~\cite{FDTferro1,a1}.
These features make the FD plot rather similar to the one obtained in
$d=1$, see Fig.~\ref{fig:spinpowerlimit} (left).
 
\begin{figure} 
\psfig{file=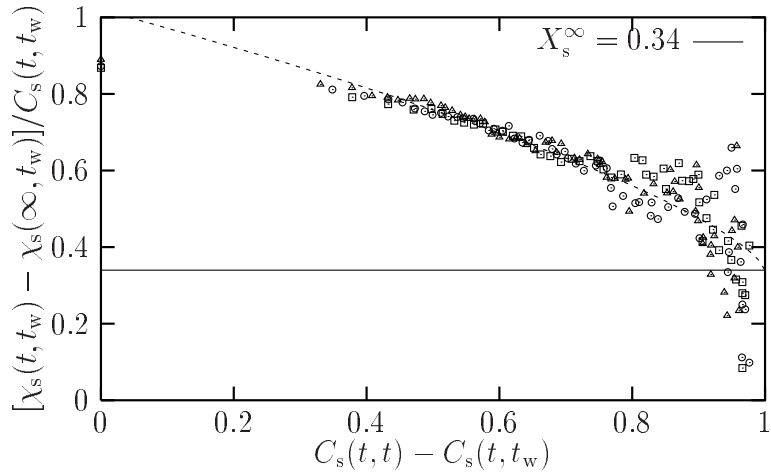,width=11cm} 
\caption{Tentative extrapolation of the infinite time slope of the FD
plots of Fig.~\protect\ref{2dspin}. The lines are only suggestive,
indicating that an asymptotic FDR of $\Xs^\infty = 0.34$ is compatible
with the data in the regime $\cors(t,t)- \cors(t,\tw) \approx 1$. The
different symbols have the same meaning as in Fig.~\ref{2dspin}.}
\label{xs} 
\end{figure} 
 
The infinite time value for the slope of the FD plot for the 2$d$
Ising model was estimated in Ref.~\cite{golu2} as
$\Xs^\infty=0.26$. We recognize from Fig.~\ref{2dspin} that the
crossover from $X=1$ to $\Xs^\infty <1$ takes place over a very small
range of the correlator, and that a precise determination of the
infinite-time value of the FDR is difficult.  A tentative numerical
extrapolation is shown Fig.~\ref{xs}, where the quantity
$(\chis(t,\tw) - \chis(\infty,\tw)) / \cors(t,\tw)$ is plotted against
$\cors(t,t) - \cors(t,\tw)$; as the abscissa approaches 1 (i.e.\ for
large time differences), the ordinate should converge to
$\Xs^\infty$. The figure shows that the value $\Xs^\infty = 0.34$ is
compatible with the data, but even though we use larger waiting times
than in Ref.~\cite{golu2} there is substantial scatter in the
points. However, we have more precise estimates of $\Xs^\infty$ to
guide us, as we now describe.
 
\begin{figure} 
\psfig{file=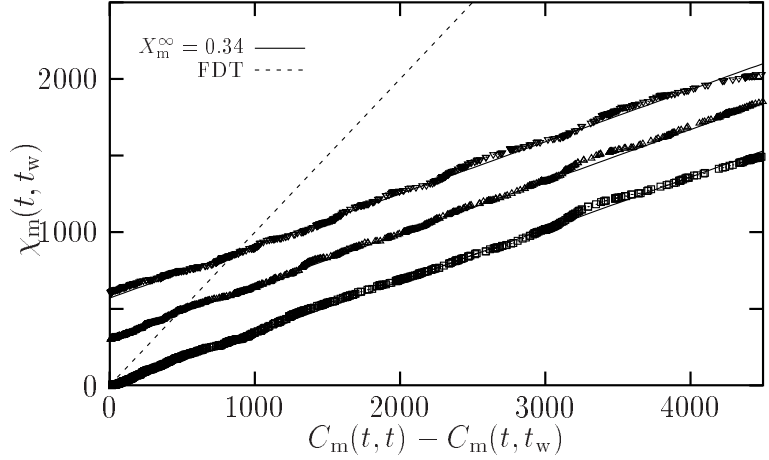,width=11cm} 
\caption{FD plots of correlation and susceptibility for 
the magnetization. The three curves 
are for waiting times $\tw=46$, 193 and 720 (bottom to top). The curves 
for $\tw=193$ and $\tw=720$ have been shifted vertically for clarity,  
since they would otherwise overlap with the curve for $\tw=46$; the 
unshifted curves all pass through the origin as they should. 
The dashed line is the equilibrium FDT. 
The full lines have slope $\Xm^\infty=0.34$.} 
\label{2dmag} 
\end{figure} 
The study of the model in $d=1$ in Sec.~\ref{section1d} showed that
the crossover from $X=1$ to $\Xs^\infty$ for spin functions reflected
the different dynamics of large and small wavevectors which for the
$d=2$ case would be defined according to $k \xi(\tw) \lessgtr 1$.
The dynamical behaviour of the small wavevectors was governed by the
asymptotic FDR $\Xs^\infty$. This suggests that a much simpler
measurement of $\Xs^\infty$ should be possible by focusing on the $k
\to 0$ limit, i.e.\ by measuring the correlation and susceptibility of
the magnetization density $m(t)$.
The resulting FD plot is reported in Fig.~\ref{2dmag}.  As for the
$1d$ case, a very simple result is obtained, with the FD plot
extremely well fitted by a simple straight line.  The fit holds for
several decades of time, $\tw < t < t_{\rm simu}$, for each waiting
time $\tw$ that we have considered, providing strong evidence that
$\Xm(t,\tw) = \Xm^\infty$ at all times.  Furthermore, the slopes of
the three curves in Fig.~\ref{2dmag} are very close to one another,
and this allows us to report the value
\begin{equation} 
\Xm^\infty = 0.340 \pm 0.005. 
\end{equation} 
This is the value we used to fit the data for the incoherent spin
functions in Figs.~\ref{2dspin} and~\ref{xs}, demonstrating that the
data are consistent with the equality $\Xs^\infty = \Xm^\infty$.  This
is somewhat different from the value reported in Ref.~\cite{golu2},
but we believe that our measurement from the magnetization is much
more reliable than the extrapolation of the incoherent spin functions,
as explained above.  We note also that this value is in extremely good
agreement with the two-loop expansion value reported in
Ref.~\cite{gambas1}.  However, unlike the $1d$ case, we do not have a
simple physical argument to explain the actual numerical value.
 
\subsection{Defect observables} 
 
\begin{figure} 
\psfig{file=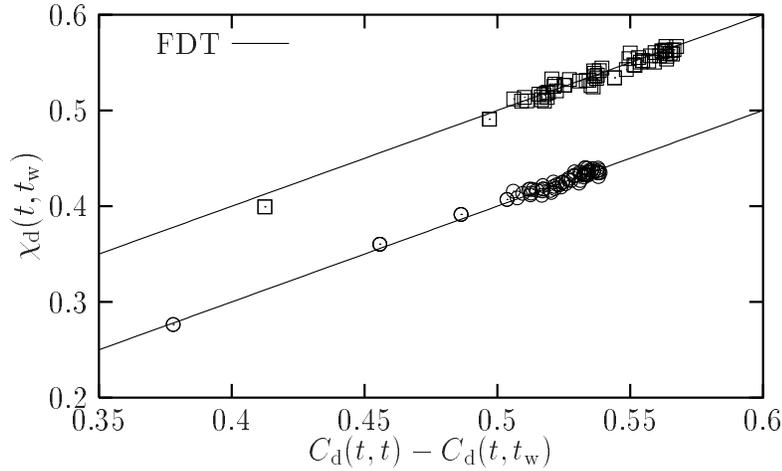,width=11cm} 
\caption{FD plots for defect autocorrelation and susceptibility. 
Data for two waiting times, $\tw=43$ (squares) and $\tw=179$ 
(circles), are presented.  
The second one has been vertically shifted for clarity. The full lines 
represent the equilibrium FDT.} 
\label{2dlink} 
\end{figure} 
We now turn to defect observables.  The simplest functions to consider
are the defect autocorrelation function and the conjugate
susceptibility. These quantities have been studied recently for
kinetically constrained Ising models (in particular the
Fredrickson-Andersen model in $1d$), where they were shown to give
rise to simple FD plots~\cite{juanpe1}.  We present the corresponding
FD plot for the $2d$ Ising model in Fig.~\ref{2dlink}. Again an
apparently very simple result is obtained, with the FD plot very well
fitted by the equilibrium straight line with $X=1$.  This is an
unexpected result, since the system is far from equilibrium as was
demonstrated by the study of spin observables in the previous
section. It could also be taken to imply, as in Ref.~\cite{juanpe1},
that the asymptotic value of the FDR associated with the defects has
the equilibrium value $\Xd^\infty = 1$.
 
Our above study of the $1d$ model again clarifies the
situation. There, we found that the incoherent dynamical functions of
the defects exhibited a crossover from equilibrium to non-equilibrium
behaviour, but that the non-equilibrium part was barely visible in an
FD plot since the crossover occurs when correlators have already
decayed to very small values. This suggests that the apparent
equilibrium behaviour observed in simulations for the $2d$ Ising and
$1d$ Fredrickson-Andersen models is simply a good approximation to
numerical data, but may miss non-trivial FD relations at large times
due to limitations in the numerical analysis.  However, as for the
spin observables, the solution to this problem is straightforward and
consists in focusing on the $k \to 0$ limit. We thus investigate next
the coherent functions for the defects which are the autocorrelation and
susceptibility for the energy density.
 
\begin{figure} 
\psfig{file=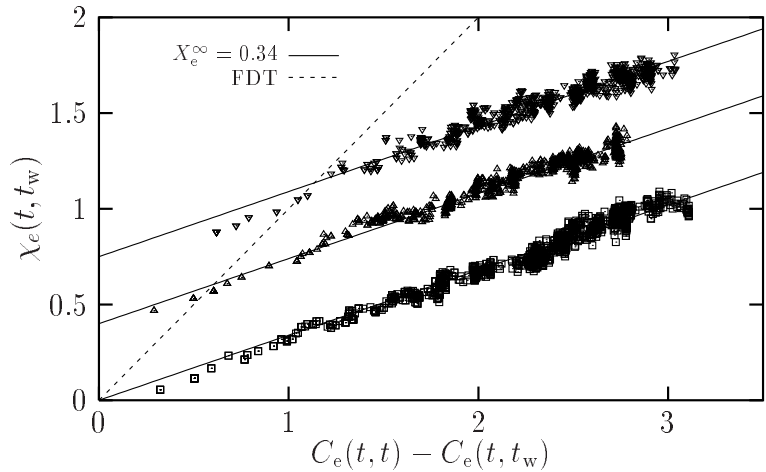,width=11cm} 
\caption{FD plots for the energy density. 
The three curves are for 
$\tw=80$, $193$, $464$ (bottom to top). 
The curves for $\tw=193$ and $\tw=464$  
have been vertically shifted for clarity, and would otherwise again 
pass through the origin. 
The dashed curve is the equilibrium FDT; the full lines have slope 
$\Xe^\infty = 0.34$.} 
\label{2dene} 
\end{figure} 
The resulting FD plot for the energy density is shown in
Fig.~\ref{2dene}.  As for the magnetization, very good fits by pure
straight lines are obtained, implying the equality $\Xe(t,\tw) =
\Xe^\infty$.  Note, however, that these plots have more noise than the
ones for the magnetization. This is due to the fact that the abscissa
now involves a genuine connected correlator, in which the nonzero
average of the energy density needs to be subtracted off.
Nonetheless, the slopes of the FD plots in Fig.~\ref{2dene} are very
close to one another and give the result
\begin{equation} 
\Xe^\infty = 0.33 \pm 0.02. 
\end{equation} 
An important outcome of this paper is that this value is compatible,
within error bars, with the value reported above for the asymptotic
FDT for the magnetization and the spins. This strongly suggests that
the various infinite time FDRs that we have measured in the $2d$ Ising
model are all equal, and that the critical point of the $2d$ Ising
model is described by a single new universal quantity
\begin{equation} 
\Xs^\infty = \Xm^\infty = \Xd^\infty = \Xe^\infty \equiv X^\infty  
\approx 0.340. 
\end{equation}

\section{Summary and discussion} 
\label{summary} 

In this paper we have studied the relation between two-time,
multi-spin, correlation and response functions in the non-equilibrium
critical dynamics of Ising models, analytically in the $d=1$ case, and
numerically in $d=2$.  We have shown that FDRs, while 
observable-dependent, 
fall into well-defined classes, which are qualitatively
similar to those observed in various glassy systems.  All FDT
violations can be understood by considering separately the
contributions from large wavevectors, which are at quasi-equilibrium
and obey FDT, and from small wavevectors where a generalized FDT holds
with a non-trivial fluctuation-dissipation ratio $\Xin=X(k \to 0)$.
In $d=1$, we find through exact calculations $\Xin = \frac{1}{2}$ for
spin observables and $\Xin = 0$ for defect observables.  In $d=2$ we
find numerically a unique $\Xin \simeq 0.34$ for all observables.
These results suggest that the definition of an effective temperature
$\Teff = T / \Xin$ for large length scales is generically possible in
non-equilibrium critical dynamics.

Further, this work also suggests many interesting lines for future
investigation. An important question is what are the limiting FDRs in
diffusive models that are analogous to the 1$d$ Ising model but have
glassy features, 
for example the one-spin facilitated Fredrickson--Andersen model~\cite{FA}
or symmetric plaquette models~\cite{juanpe1}. 
Also, it would be interesting 
to confirm our results for the $2d$ Ising model 
by analyzing higher order correlation functions by means
of the RG techniques used in Ref.~\cite{gambas1,gambas2,gambas3}
to confirm the uniqueness of the FDR. 
This would make this function an interesting quantity to study
in more generic non-equilibrium situations such as 
driven interfaces or driven diffusive systems.

\begin{acknowledgments} 
We acknowledge financial support from \"{O}sterreichische Akademie der 
Wissenschaften and EPSRC Grant No.\ 00800822 (PM), 
Marie Curie Fellowship No HPMF-CT-2002-01927, CNRS 
and Worcester College, Oxford (LB),
EPSRC Grant No.\ GR/R83712/01 (LB and JPG), the Glasstone Fund (JPG),
and Nuffield Grant No.\ NAL/00361/G (PS). 
Numerical results were obtained on OSWELL at the Oxford Supercomputing
Center, Oxford University, UK
\end{acknowledgments}

\appendix 
\section{Modified Bessel Functions} 
\label{sec:bessel} 
 
Here we briefly summarize the main properties of modified Bessel functions  
$\In_n(x)$ that are relevant for the analysis given above. A  
comprehensive description may be found in \cite{mathbook}. For integer  
order $n$, $\In_n(x)$ has the integral representation 
\begin{equation} 
  \In_n(x) = \int\limits_0^\pi \frac{\upd \varphi}{\pi} \cos n\varphi \; 
  \me^{x \cos \varphi}, \label{equ:Inintegral} 
\end{equation} 
from which the functional relations  
\begin{equation} 
  \frac{\partial}{\partial x}\In_n(x) = \frac{1}{2}  
    [\In_{n-1}+\In_{n+1}](x) \quad \mbox{and} \quad  
  \frac{2n}{x} \In_n(x) = [\In_{n-1}-\In_{n+1}](x)  
    \label{equ:Inprop} 
\end{equation} 
follow immediately. In particular it is clear from (\ref{equ:Inintegral})  
that $\In_{-n}(x)=\In_n(x)$ and $\In_n(-x)=(-1)^n \In_n(x)$. The aging  
expansions of our results are based on the asymptotic formula 
\begin{equation} 
  \In_n(x) = \frac{\me^x}{\sqrt{2\pi x}} \left( 1+\frac{1-4n^2}{8x} +  
  \mathcal{O}\left(\frac{1}{x^2}\right) \right), \label{equ:Inasym} 
\end{equation} 
which applies in the limit of large arguments $x$ for fixed  
order $n$. For the derivation of the Fourier transforms of multi-spin  
correlation and response functions we use  
\begin{eqnarray} 
  \sum_n \me^{-i n x} \In_n(a) & = & \me^{a \cos x}, \label{equ:Insum1}\\ 
  \sum_n \me^{-i n k} \In_{n-m}(a) \In_{n+m}(a) & = & \In_{2m}\left(2a\cos 
    \frac{k}{2}\right) , \label{equ:Insum2} \\ 
  \sum_n \me^{-i n k} \In_n(a) (\In_{n+m}(b)+\In_{n-m}(b)) & = & 2  
    \mathrm{T}_m\left(\frac{b+a\cos k}{A}\right) \In_m(A) , 
  \label{equ:Insum3}  
\end{eqnarray} 
where (\ref{equ:Insum2}), (\ref{equ:Insum3}) follow from  
(\ref{equ:Inintegral}), the well known identity (\ref{equ:Insum1}) and  
trigonometric relations. In (\ref{equ:Insum3}) $0<a \leq b$ is required,   
$A=\sqrt{a^2+b^2+2ab\cos k}$ and the $\mathrm{T}_n(x)=\cos(n \arccos x)$  
are Chebyshev polynomials of degree $n$ in $x$.

\section{Fourier Transforms} 
\label{sec:fourier} 
 
The Fourier transforms of spin correlation and response functions  
(\ref{equ:corsn}), (\ref{equ:chisn}) follow immediately when using  
(\ref{equ:Insum1}):
\begin{eqnarray} 
C(k;t,\tw) & = & \me^{-(t+\tw)(1-\cos k)} \Bigg\{ 1 +  
  \int\limits_0^{2\tw} \upd\tau \;\; \me^{-\tau \cos k}  
  [\In_0+\In_1](\tau) \Bigg\} , \label{equ:corsfourier} \\ 
\chi(k;t,\tw) & = & \frac{1}{2} \int\limits_{\tw}^t  
  \upd\tau \;\; \me^{-(t-\tau)(1-\cos k)} \; \me^{-2 \tau} \left[\In_0 + 
  2 \In_1 + \In_2 \right] (2\tau) . \label{equ:chisfourier} 
\end{eqnarray} 
For defect correlations, however, a direct transformation of  
(\ref{equ:cordn}) yields a rather intractable expression. Therefore  
we first rewrite (\ref{equ:cordn}) using the identity (which can be  
verified by differentiation)  
\begin{equation*} 
  \me^{-(t-\tw)} \In_n(t-\tw) = \me^{-(t+\tw)} \In_n(t+\tw) -  
    \int\limits_{t-\tw}^{t+\tw} \upd\tau \, \frac{1}{2} \me^{-\tau} [ 
    \In_{n-1}-2\In_n+\In_{n+1}](\tau) , 
\end{equation*} 
as 
\begin{align*} 
  & \cord[n](t,\tw) = \me^{-2 (t+\tw)}  
    [ \In_n^2-\In_{n-1}\In_{n+1} ](t+\tw) +  
    \frac{1}{2} \int\limits_{t-\tw}^{t+\tw}  
    \upd\tau \, \me^{-(t+\tw+\tau)} \, \times\\ 
  & \quad \Big\{ [\In_{n-1}-\In_{n+1}](\tau) \;  
    [\In_{n-1}-\In_{n+1}](t+\tw)  
  - [\In_{n-1}-2\In_n+\In_{n+1}](\tau) \;  
    [\In_{n-1}+2\In_n+\In_{n+1}](t+\tw) \Big\} . \nonumber  
\end{align*} 
Now, utilizing (\ref{equ:Inprop}) and expressing factors of $n$ as  
derivatives w.r.t.\ $k$, the Fourier series for $C(k;t,\tw)$ may be  
written in the form: 
\begin{align*} 
  & C(k;t,\tw) = \me^{-2 (t+\tw)} \sum_n \me^{-i n k}  
    [\In_n^2 - \In_{n-1} \In_{n+1}](t+\tw) - \frac{1}{2}  
    \int\limits_{t-\tw}^{t+\tw} \upd\tau \, \me^{-(t+\tw+\tau)} \, \times \\ 
  & \quad  \left\{ \frac{4}{\tau (t+\tw)} \frac{\partial^2}{\partial 
    k^2} \sum_n  
    \me^{-i n k} \In_n(\tau) \In_n(t+\tw) + 2 \left( 
    \frac{\partial}{\partial \tau}-1 \right) \sum_n \me^{-i n k}  
    \In_n(\tau) [\In_{n-1}+2\In_n+\In_{n+1}](t+\tw) \right\} . \nonumber 
\end{align*} 
All summations in this expression can be evaluated via (\ref{equ:Insum2}),  
(\ref{equ:Insum3}). Some fairly complicated algebra is required 
to simplify the resulting expression, but finally one obtains 
the compact result  
\begin{eqnarray} 
  C(k;t,\tw) & = & { \me^{-2 (t+\tw)} \frac{\In_1(2 (t+\tw)  
    \cos\frac{k}{2})}{(t+\tw) \cos\frac{k}{2}} } \label{equ:cordfourier} \\ 
  & + & { 4\int\limits_0^{\tw} \upd\tau \me^{-2 (t+\tau)}  
    \Bigg[ \frac{1}{A} \In_1(2 A) + 2 \frac{\tw-\tau}{A} \sin^2
    \left(\frac{k}{2}\right)  
    \left( \In_1(2 A) + \frac{\tw-\tau}{A} \In_2(2 A) \right) \Bigg] } ,
    \nonumber 
\end{eqnarray} 
where $A=\sqrt{(t+\tau)^2\cos^2({k}/{2}) + (\tw-\tau)^2\sin^2 
({k}/{2})}$. Eq.~(\ref{equ:cordfourier}) is the most convenient  
representation for $C(k;t,\tw)$, both for numerical and 
analytical purposes. The calculation of the Fourier transform of the defect  
susceptibility (\ref{equ:chidn}) is comparatively easy; from  
(\ref{equ:Insum2}), (\ref{equ:Insum3}) one finds 
\begin{equation} 
  \chi(k;t,\tw)=2 \me^{-2t} \left\{  
  [\In_0+\In_1](2t)-\In_0(2 A)- 
  \frac{t\cos^2({k}/{2})+\tw\sin^2({k}/{2})}{A} \In_1(2 A) \right\} ,
  \label{equ:chidfourier} 
\end{equation} 
with $A=\sqrt{t^2\cos^2({k}/{2})+\tw^2\sin^2({k}/{2})}$.

\section{Power-law Covariances} 
\label{sec:powerQ} 
 
Here we show that the covariances $q_{\power,n}$ given in  
(\ref{equ:powerQ}) follow a power-law as $|n| \to \infty$ and  
establish the link $q_{\power,n}=\mathcal{F}^{-1}\{q_{\power}(k)\}$.  
Let us first focus on the Fourier integral (\ref{equ:fourier}) which  
-- since $q_\power(k)$ is even in $k$ -- may be written as 
\begin{equation} 
  q_{\power,n}=\frac{\Gamma^2\left(\frac{1+\alpha}{2}\right)} 
  {2^{1-\alpha} \Gamma(\alpha)} \int\limits_0^{2\pi} \frac{\upd k}{2\pi} \,  
  \left(\sin\frac{k}{2}\right)^{\alpha-1} \cos ( n k) ,
  \label{equ:Qintegral} 
\end{equation} 
where $0<\alpha<1$ as before. The simple substitution $x=k/2$ yields 
the solvable integral~\cite{mathbook} 
\begin{equation} 
  \int_0^\pi \frac{\upd x}{\pi} \, (\sin x)^{\alpha-1} \cos 2 n x =  
  \frac{(-1)^n}{\alpha 2^{\alpha-1} \mathrm{B}  
  \left( \frac{1+\alpha}{2}+n,\frac{1+\alpha}{2}-n \right) } .
  \label{equ:Qsolution} 
\end{equation} 
Using the functional relation \cite{mathbook} $\mathrm{B}(x,y)=\Gamma(x)  
\Gamma(y) / \Gamma(x+y)$ for the Beta function $\mathrm{B}(x,y)$ and  
simplifying the remaining expression yields the result for  
$q_{\power,n}$ given in  
(\ref{equ:powerQ}). Now we turn to the asymptotic behaviour of  
$q_{\power,n}$ as $|n| \to \infty$. For $n \geq 1$ we may rewrite  
$q_{\power,n}$, using $\Gamma(x) \Gamma(1-x) = \pi/\sin \pi x$ and  
$\Gamma(x+1) = x \Gamma(x)$, in the form  
\begin{equation} 
  q_{\power,n} = \prod_{k=0}^{n-1} \frac{1-\alpha+2k}{1+\alpha+2k} .
  \label{equ:Qproduct} 
\end{equation} 
It is obvious from (\ref{equ:Qproduct}) that $q_{\power,n}$ is monotonically  
decreasing and vanishes for $n \to \infty$ as long as $\alpha > 0$.  
It is equally clear that $q_{\power,n}=1$ for  
$\alpha \to 0$ and $q_{\power,n}=\delta_{n,0}$ as $\alpha \to 1$ (since  
$q_{\power,n}$ is even in $n$ and $q_{\power,0}=1 \, \forall \alpha$). 
In order to understand the asymptotic behaviour of $q_{\power,n}$ we  
take the logarithm of (\ref{equ:Qproduct}) and use the bounds  
\begin{equation} 
  \int\limits_0^n \upd k \, a_k \leq \sum_{k=0}^{n-1} a_k \leq a_0 +  
  \int\limits_0^{n-1} \upd k \, a_k ,
  \label{equ:bounds} 
\end{equation} 
which hold for any non-increasing function $a_k$. For the case  
at hand the integrals can be solved easily. Exponentiating the result,  
multiplying by  $n^\alpha$ and taking the limit $n \to \infty$ then  
gives  
\begin{equation} 
  (2 \me)^{-\alpha} 
  \sqrt{\frac{(1-\alpha)^{1-\alpha}}{(1+\alpha)^{1+\alpha}}}  
  \leq \lim_{n \to \infty} n^\alpha q_{\power,n} \leq  
  (2 \me)^{-\alpha} 
  \sqrt{\frac{(1+\alpha)^{1-\alpha}}{(1-\alpha)^{1+\alpha}}}, 
\end{equation} 
which implies that there exists a finite constant $c$ such that  
$q_{\power,n} \sim c \, n^{-\alpha}$ for $0<\alpha<1$.


\begin{thebibliography}{99}
\bibitem{defectformula} For $t=\tw$ the integral in (\ref{equ:corsn})
is of the form of a Laplace convolution. A Laplace transform indeed
allows one to simplify this expression and rewrite it explicitly in
terms of modified Bessel functions.
\end{thebibliography}
\end{document}